\documentclass[review]{elsarticle}

\usepackage{hyperref}
\usepackage[numbered]{bookmark}
\usepackage{soul}
\usepackage{booktabs}
\usepackage{multirow}
\usepackage{float}
\usepackage{url}
\usepackage[small,it]{caption}
\usepackage{tcolorbox}
\usepackage{amsmath,amssymb,amsfonts}
\usepackage{algorithmic}
\usepackage{graphicx}
\usepackage{textcomp}
\usepackage{xcolor}
\usepackage{tabularx} 
\usepackage{balance}
\usepackage{enumitem}

\journal{Journal of Systems and Software}

\newcommand{\RQA}{What are the most frequent human-annotated grammar patterns and what are the semantics of these patterns?}

\newcommand{\RQB}{How accurately do the chosen taggers annotate grammar patterns and individual tags?}

\newcommand{\RQC}{Are there other grammar patterns that are dissimilar from the most frequent in our data, but still present in multiple systems?}

\newcommand{\RQD}{Do grammar patterns or tagger accuracy differ across programming languages?}

\begin{document}

\begin{frontmatter}
\title{On the Generation, Structure, and Semantics of Grammar Patterns in Source Code Identifiers}
\cortext[mycorrespondingauthor]{Corresponding author}
\author{Christian D. Newman\fnref{ritfoot}\corref{mycorrespondingauthor}}
\ead{cnewman@se.rit.edu}
\author{Reem S. AlSuhaibani\fnref{reemfoot}}
\ead{ralsuhai@kent.edu}
\author{Michael J. Decker\fnref{bgsufoot}}
\ead{mdecke@bgsu.edu}
\author{Anthony Peruma\fnref{ritfoot}} 
\ead{ axp6201@rit.edu}
\author{Dishant Kaushik\fnref{ritfoot}} 
\ead{dkaushik95@gmail.com}
\author{Mohamed Wiem Mkaouer\fnref{ritfoot}} 
\ead{mwmvse@rit.edu}
\author{Emily Hill\fnref{drewfoot}}
\ead{emhill@drew.edu}
\fntext[ritfoot]{Rochester Institute of Technology, Rochester, NY, USA}
\fntext[reemfoot]{Kent State University, Kent, OH, USA, Prince Sultan University, Riyadh, KSA}
\fntext[bgsufoot]{Bowling Green State University, Bowling Green, OH, USA}
\fntext[drewfoot]{Drew University, Madison, NJ, USA}

\begin{abstract}
Identifiers make up a majority of the text in code. They are one of the most basic mediums through which developers describe the code they create and understand the code that others create. Therefore, understanding the patterns latent in identifier naming practices and how accurately we are able to automatically model these patterns is vital if researchers are to support developers and automated analysis approaches in comprehending and creating identifiers correctly and optimally. This paper investigates identifiers by studying sequences of part-of-speech annotations, referred to as grammar patterns. This work advances our understanding of these patterns and our ability to model them by 1) establishing common naming patterns in different types of identifiers, such as class and attribute names; 2) analyzing how different patterns influence comprehension; and 3) studying the accuracy of state-of-the-art techniques for part-of-speech annotations, which are vital in automatically modeling identifier naming patterns, in order to establish their limits and paths toward improvement. To do this, we manually annotate a dataset of 1,335 identifiers from 20 open-source systems and use this dataset to study naming patterns, semantics, and tagger accuracy.
\end{abstract}

\end{frontmatter}

% \section{The Elsevier article class}

% \paragraph{Installation} If the document class \emph{elsarticle} is not available on your computer, you can download and install the system package \emph{texlive-publishers} (Linux) or install the \LaTeX\ package \emph{elsarticle} using the package manager of your \TeX\ installation, which is typically \TeX\ Live or Mik\TeX.

% \paragraph{Usage} Once the package is properly installed, you can use the document class \emph{elsarticle} to create a manuscript. Please make sure that your manuscript follows the guidelines in the Guide for Authors of the relevant journal. It is not necessary to typeset your manuscript in exactly the same way as an article, unless you are submitting to a camera-ready copy (CRC) journal.

% \paragraph{Functionality} The Elsevier article class is based on the standard article class and supports almost all of the functionality of that class. In addition, it features commands and options to format the
% \begin{itemize}
% \item document style
% \item baselineskip
% \item front matter
% \item keywords and MSC codes
% \item theorems, definitions and proofs
% \item lables of enumerations
% \item citation style and labeling.
% \end{itemize}

\section{Introduction}
\label{introduction}
Currently, developers spend a significant amount of time comprehending code \cite{Corbi1989, Martin:2008}; 10 times the amount they spend writing it by some estimates \cite{Martin:2008}. Studying comprehension will lead to ways that not only increase the ability of developers to be productive, but also increase the accessibility of software development (e.g., by supporting programmers that prefer top-down or bottom-up comprehension styles \cite{vonMayrhauser:1997,Fisher:2006}) as a field. One of the primary ways a developer comprehends code is by reading identifier names which make up, on average, about 70\% of the characters found in a body of code \cite{Deissenbock:2005}. 

Recent studies show how identifier names impact comprehension \cite{Schankin:2018, Binkley2006, Hofmeister:2017, butler2010exploring, Takang1996}, others show that normalizing identifier names helps both developers and research tools which leverage identifiers \cite{Binkley:2018, newmanabbrev}. Thus, many research projects try to improve identifier naming practices. For example, prior research predicts words that should appear in an identifier name \cite{Kashiwabara:2014, Allamanis:2015, Liu:2019, Abebe:2013}; combines Natural Language (NL) and static analysis to analyze or suggest changes to identifiers \cite{Arnaoudova:2013, Arnaoudova:2014,Perumascam, Host:2009, Abebe:2013, Peruma:2018:EIW:3242163.3242169, Binkley:2011, Gupta:2013, Hill:2011}; and groups identifier names by static role in the code \cite{Dragan:2006, Alsuhaibani:2015, Newman:2017}. 

One challenge to studying identifiers is the difficulty in understanding how to map the meaning of natural language phrases to the behavior of the code. For example, when a developer names a method, the name should reflect the behavior of the method such that another developer can understand what the method does without the need to read the method body instruction set. Understanding this connection between name and behavior presents challenges for humans and tools; both of which use this relationship to comprehend, generate, or critique code. A second challenge lies in the natural language analysis techniques themselves, many of which are not trained to be applied to software \cite{Binkley:2018}, which introduces significant threats \cite{Jongeling:2017}. Addressing these problems is vital to improving the developer experience and augmenting tools which leverage natural language.

One of the most popular methods for measuring the natural language semantics of identifier names is part-of-speech (POS) tagging. While some work has been done studying part-of-speech tag sequences (i.e., grammar patterns) in identifier names \cite{Host:2009, Gupta:2013, Shepherd:2009, butler:2011, butler:2015,Liblit06cognitiveperspectives}, these prior works either focus on a specific type of identifier, typically method or class names; or discuss grammar patterns at a conceptual level without showing concrete examples of what they look like in the wild, where they can be messy, incomplete, ambiguous, or provide unique insights about how developers express themselves. In this paper, we begin addressing these issues. We create a dataset of 1,335 manually-annotated (i.e., POS tagged) identifiers across five identifier categories: class, function, declaration-statement  (i.e., global or function-local variables), parameter, and attribute names. We use this dataset to study and show concrete grammar patterns as they are found in their natural environments.

The goal of this paper is to study the structure, semantics, diversity, and generation of grammar patterns. We accomplish this by 1) establishing and exploring the common and diverse grammar pattern structures found in identifiers. 2) Using these structures to investigate the accuracy, strengths, and weaknesses of approaches which generate grammar patterns with an eye toward establishing and improving their current ability. Finally, 3) leveraging the grammar patterns we discover to discuss the ways in which words, as part of a larger identifier, work together to convey information to the developer. We answer the following Research Questions (RQs):

\textbf{RQ1: \RQA}  This question explores the top 5 frequent patterns generated by the human annotators and discusses what bearing these patterns have on comprehending the connection between identifiers and code semantics/behavior.

\textbf{RQ2: \RQB} This question explores the accuracy of the part-of-speech tagger annotations versus human annotations. We determine which patterns and part-of-speech tags are most often incorrectly generated by tools.

\textbf{RQ3: \RQC} This question explores patterns which are not as frequent as those discussed in RQ1. We manually pick a set of patterns that are structurally dissimilar from the top 5 from RQ1, but still appear in multiple systems within the dataset. Consequently, we identify unique patterns which are not as regularly observed as our top 5 patterns, but are repeatedly represented in the dataset and important to discuss. This question addresses the diversity of patterns within the dataset.

\textbf{RQ4: \RQD} This question explores how grammar patterns compare when separated by programming language. We determine how C/C++ and Java grammar patterns are structurally similar and dissimilar from one another. We also analyze tagger accuracy when we split grammar patterns by programming language.

We also discuss the ways which we will use this data in our future work. The results of this study:
\begin{itemize} \setlength\itemsep{0em}
    \item Increase our understanding of how different grammar patterns convey information.
    \item Provide a list of patterns which may be used to both understand naming practices and identify further patterns requiring further study.
    \item Highlight the accuracy of, and ways to improve, part-of-speech taggers.
    \item Highlight the prevalence, and usage, of part-of-speech annotations which are found in source code.
\end{itemize}

This paper is organized as follows, Section \ref{grammarpatterndef} gives the necessary background related to grammar pattern generation. Our methodology is detailed in Section \ref{methodology}. Experiments, driven by answering our research questions, are in Section \ref{evaluation}. Section \ref{threats} enumerates all the threats to the validity of our experiments. Section \ref{sec:related} discusses studies related to our work, before further discussing our findings in Section \ref{Discussion+conclusion} and concluding in Section \ref{Conclusion}.

\section{Definitions \& Grammar Pattern Generation}
\label{grammarpatterndef}
In this paper, we use grammar patterns to understand the relationship between groups of identifiers. A \textit{grammar pattern} is the sequence of part-of-speech tags (also referred to as annotations) assigned to individual words within an identifier. For example, for an identifier called GetUserToken, we assign a grammar pattern by splitting the identifier into its three constituent words: Get, User, and Token. We then run the split-sequence (i.e., Get User Token) through a part-of-speech tagger to get the grammar pattern: Verb Adjective Noun. Notice this grammar pattern is not unique to this identifier, but is shared with many potential identifiers that use similar words. Thus, we can relate identifiers that contain different words to one another using their grammar pattern; GetUserToken, RunUserQuery, and WriteAccessToken share the same grammar pattern and, while they do not express the exact same semantics, there are similarities in their semantics which their grammar patterns reveal. Specifically, a verb (get, run, write) applied to a noun (token, query) with a specific role/context (user, access).

\begin{table}[]
\centering
\caption{Examples of grammar patterns}
\label{Table:GrammarPatternExample}
\begin{tabular}{@{}ll@{}}
\toprule
\multicolumn{1}{c}{\textbf{Identifier Example}} & \multicolumn{1}{c}{\textbf{Grammar  Pattern}} \\ \midrule
1. GList* \textbf{tile list head} = NULL; & adjective adjective noun \\
2. GList* \textbf{tile list tail} = NULL; & adjective adjective noun \\
3. Gulong \textbf{max tile size} = 0; & adjective adjective noun \\
4. GimpWireMessage \textbf{msg}; & noun \\
5. \textbf{g list remove link} (tile list head, list) & preamble noun verb noun \\
6. \textbf{g list last} (list) & preamble adjective noun \\
7. \textbf{g assert} (tile\_list\_head != tile\_list\_tail); & preamble verb \\
\bottomrule
\end{tabular}
\end{table}
\begin{table}[]
\setlength{\tabcolsep}{1.5pt}
\centering
\caption{Part-of-speech categories used in study}
\label{tab:posusedtable}
\begin{tabular}{@{}lll@{}}
\toprule
\textbf{Abbreviation} & \textbf{Expanded Form} & \textbf{Examples} \\ \midrule
N & noun & \begin{tabular}[c]{@{}l@{}}Disneyland, shoe, faucet,\\ mother, bedroom\end{tabular} \\ \midrule
DT & determiner & \begin{tabular}[c]{@{}l@{}}the, this, that, these, those, which\end{tabular} \\ \midrule
CJ & conjunction & and, for, nor, but, or, yet, so \\ \midrule
P & preposition & \begin{tabular}[c]{@{}l@{}}behind, in front of, at, under,\\ beside, above, beneath, despite\end{tabular} \\ \midrule
NPL & noun plural & \begin{tabular}[c]{@{}l@{}}Streets, cities, cars, people,\\ lists, items, elements.\end{tabular} \\ \midrule
NM & noun modifier & \begin{tabular}[c]{@{}l@{}}red, cold, hot, scary, beautiful,\\ happy, faster, small\end{tabular} \\ \midrule
V & verb & Run, jump, drive, spin \\ \midrule
VM & verb modifier (adverb) & \begin{tabular}[c]{@{}l@{}}Very, loudly, seriously,\\ impatiently, badly\end{tabular} \\ \midrule
PR & pronoun & \begin{tabular}[c]{@{}l@{}}she, he, her, him, it, we,\\ us, they, them, I, me, you\end{tabular} \\ \midrule
D & digit & 1, 2, 10, 4.12, 0xAF \\ \midrule
PRE & preamble* & Gimp, GLEW, GL, G \\ \bottomrule
\end{tabular}
\end{table}
\subsection{Generating Grammar Patterns}
To generate grammar patterns for identifiers in source code, we require a part-of-speech tagger and a way to split identifiers into their constituent terms. To split terms in an identifier, we use Spiral \cite{HuckaSpiral}; an open-source tool combining several splitting approaches into one Python package. From this package, we used heuristic splitting and manually corrected mistakes made by the splitter in our manually-annotated set, which we discuss in the next section. We generate grammar patterns by running each individual tagger on each identifier after applying our splitting function. Additionally, we give examples of grammar patterns in Table~\ref{Table:GrammarPatternExample}, which shows a set of identifiers on the left and the corresponding grammar pattern on the right.

Since part of the goal of this paper is to study POS taggers, we use multiple taggers on our identifier dataset; Posse \cite{Gupta:2013}, Swum \cite{HillSwum:2010}, and Stanford \cite{Toutanova:StanfordTagger} \footnote{Version: 3.9.2, taggermodel: english-bidirectional-distsim.tagger, jdk version: openJDK 11.0.7}. Posse and Swum are part-of-speech taggers created specifically to be run on software identifiers; they are trained to deal with the specialized context in which identifiers appear. Both Posse and Swum take advantage of static analysis to provide annotations. For example, they will look at the return type of a function to determine whether the word \textit{set} is a noun or a verb. Additionally, they are both aware of common naming structures in identifier names. For example, methods are more likely to contain a verb in certain positions within their name (e.g., at the beginning) \cite{Gupta:2013,HillSwum:2010}. They leverage this information to help determine what POS to assign different words. Stanford is a popular POS tagger for general natural language (e.g., English) text. For our study, Stanford provides a baseline; it is not specialized for source code but is reasonably accurate on method names \cite{Olney2016}.

This paper uses the part-of-speech tags given in Table~\ref{tab:posusedtable}. Note, in this table, the \textit{preamble} category, which does not exist in general natural language part-of-speech tagging approaches. A Preamble is an abbreviation which does one of the following: 
\begin{enumerate}[nolistsep]
    \item Namespaces an identifier without augmenting the reader's understanding of its behavior (e.g., XML in XML\_Reader is not a preamble)
    \item Provides language-specific metadata about an identifier (e.g., identifies pointers or member variables)
    \item Highlights an identifier's type. When a preamble is highlighting an identifier's type, the type's inclusion must not add any new information to the identifier name.
\end{enumerate}
For example, given an identifier \textit{float* fPtr}, 'f' does not add any information about the identifier's role within the system, but reminds the developer that it has a type 'float'. However, given an identifier \textit{char* sPtr}, 's' informs the developer that this is a c-string as opposed to a pointer to some other type of character array; 's' would not be considered a preamble under this definition. Additionally, some developers will put \textit{p\_} in front of pointer variables or m\_ in front of variables that are members of a class; these are due to naming conventions and/or Hungarian notation \cite{butler2010exploring,butler:2015,hillamap, hungariannotation}. In the GIMP and GLEW open-source projects, GIMP and G\_ are preambles to many variables, as seen in the Gimp example in Table~\ref{Table:GrammarPatternExample}. Intuitively, the reason for identifying preambles in an identifier is because they do not provide any information with respect to the identifier's role within the system's domain. Instead, they provide one of the three types of information above. For this reason, when analyzing identifiers, tools should be able to determine when a word is a preamble so that they do not make false assumptions about the word's descriptive purpose.

Another tag to note from Table \ref{tab:posusedtable} is \textit{noun modifier (NM)}, which is annotated on words that could be considered either pure adjectives or noun-adjuncts. A noun-adjunct is a word that is typically a noun but is being used as an adjective. An example of this is the word \textit{bit} in the identifier \textit{bitSet}. In this case, \textit{bit} is a noun which describes the type of \textit{set}, i.e., it is a set of bits. So we consider it a noun-adjunct. These are found in English (e.g., compound words), but generally not annotated as their own individual part-of-speech tag. In this work, we argue for the use of an individual tag for noun-adjuncts due to their ubiquity, and special role, in source code identifiers, which we discuss.

\begin{table}[]
\centering
\caption{How Penn Treebank annotations were mapped to the reduced set of annotations}
\label{tab:penntreebankmap}
\resizebox{!}{.28\paperheight}{%
\begin{tabular}{@{}c|c@{}}
\toprule
{\color[HTML]{000000} \begin{tabular}[c]{@{}c@{}}Penn Treebank\\ Annotation\end{tabular}} & {\color[HTML]{000000} \begin{tabular}[c]{@{}c@{}}Annotation Used\\ In Study\end{tabular}} \\ \midrule
{\color[HTML]{000000} Conjunction (CC)} & {\color[HTML]{000000} Conjunction (CJ)} \\ \midrule
{\color[HTML]{000000} Digit (CD)} & {\color[HTML]{000000} Digit (D)} \\ \midrule
{\color[HTML]{000000} Determiner (DT)} & {\color[HTML]{000000} Determiner (DT)} \\ \midrule
{\color[HTML]{000000} Foreign Word (FW)} & {\color[HTML]{000000} Noun (N)} \\ \midrule
{\color[HTML]{000000} Preposition (IN)} & {\color[HTML]{000000} Preposition (P)} \\ \midrule
{\color[HTML]{000000} Adjective (JJ)} & {\color[HTML]{000000} Noun Modifier (NM)} \\ \midrule
{\color[HTML]{000000} Comparative Adjective (JJR)} & {\color[HTML]{000000} Noun Modifier (NM)} \\ \midrule
{\color[HTML]{000000} Superlative Adjective (JJS)} & {\color[HTML]{000000} Noun Modifier (NM)} \\ \midrule
{\color[HTML]{000000} List Item (LS)} & {\color[HTML]{000000} Noun (N)} \\ \midrule
{\color[HTML]{000000} Modal (MD)} & {\color[HTML]{000000} Verb (V)} \\ \midrule
{\color[HTML]{000000} Noun Singular (NN)} & {\color[HTML]{000000} Noun (N)} \\ \midrule
{\color[HTML]{000000} Proper Noun (NNP)} & {\color[HTML]{000000} Noun (N)} \\ \midrule
{\color[HTML]{000000} Proper Noun Plural (NNPS)} & {\color[HTML]{000000} Noun Plural (NPL)} \\ \midrule
{\color[HTML]{000000} Noun Plural (NNS)} & {\color[HTML]{000000} Noun Plural (NPL)} \\ \midrule
{\color[HTML]{000000} Personal Pronoun (PRP)} & {\color[HTML]{000000} Pronoun (PR)} \\ \midrule
{\color[HTML]{000000} Possessive Pronoun (PRP\$)} & {\color[HTML]{000000} Pronoun (PR)} \\ \midrule
{\color[HTML]{000000} Adverb (RB)} & {\color[HTML]{000000} Verb Modifier (VM)} \\ \midrule
{\color[HTML]{000000} Comparative Adverb (RBR)} & {\color[HTML]{000000} Verb Modifier (VM)} \\ \midrule
{\color[HTML]{000000} Particle (RP)} & {\color[HTML]{000000} Verb Modifier (VM)} \\ \midrule
{\color[HTML]{000000} Symbol (SYM)} & {\color[HTML]{000000} Noun (N)} \\ \midrule
{\color[HTML]{000000} To Preposition (TO)} & {\color[HTML]{000000} Preposition (P)} \\ \midrule
{\color[HTML]{000000} Verb (VB)} & {\color[HTML]{000000} Verb (V)} \\ \midrule
{\color[HTML]{000000} Verb (VBD)} & {\color[HTML]{000000} Verb or Noun Modifier (V or NM**)} \\ \midrule
{\color[HTML]{000000} Verb (VBG)} & {\color[HTML]{000000} Verb or Noun Modifier (V or NM**)} \\ \midrule
{\color[HTML]{000000} Verb (VBN)} & {\color[HTML]{000000} Verb or Noun Modifier (V or NM**)} \\ \midrule
{\color[HTML]{000000} Verb (VBP)} & {\color[HTML]{000000} Verb (V)} \\ \midrule
{\color[HTML]{000000} Verb (VBZ)} & {\color[HTML]{000000} Verb (V)} \\ \bottomrule
\end{tabular}
}
\end{table}

\begin{figure}[h]
\centering
\includegraphics[]{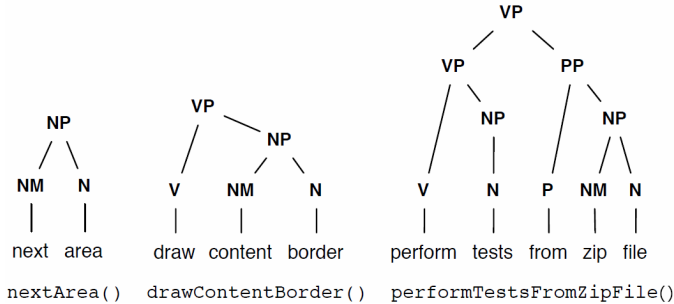}
\caption{Examples of noun, verb, and prepositional phrases}
\label{fig:example_phrasal_concepts}
\end{figure}

The tagset in Table~\ref{tab:posusedtable} is a smaller set than some standard natural language part-of-speech tagsets, such as the Penn Treebank tagset used by Stanford \cite{Toutanova:StanfordTagger}, due to the fact that Posse \cite{Gupta:2013} and Swum \cite{HillSwum:2010} use this limited set. Because Swum/Posse both rely on a more limited set, we use a manually curated mapping from the Penn Treebank set to this narrower tagset, provided in Table~\ref{tab:penntreebankmap}, which we now discuss. Many of these mappings take subcategories of various part-of-speech annotations and translate them to the broadest category. For example, proper noun $\longrightarrow$ noun and modal $\longrightarrow$ verb, where a modal is just a more specific kind of verb, and a proper noun is a more specific kind of noun.

Past tense verb (VBD), present participle verb (VBG), and past participle verb (VBN) are all used as adjectives in many situations within our data set. For example, \textit{sortedIndicesBuf}, \textit{waitingList}, and \textit{adjustedGradient} where \textit{sorted} is a past tense verb (VBD), \textit{waiting} is a present participle (VBG), and \textit{adjusted} is a past participle verb (VBN). All three of these verbs can be used as adjectives. Therefore, we process these twice for Stanford, which is the only tagger that detects these verb subcategories: once as verbs and once as adjectives to compare Stanford's accuracy under both configurations. VBZ (third-person verbs) are also sometimes used as adjectives, but based on our data this annotation is most frequently used for the words \textit{is} or \textit{equals}; typically in boolean identifiers or methods. Therefore, we treat VBZ as a verb since treating them as an adjective would generally result in lower accuracy. The same applies to VB and VBP; words which receive these annotations are typically verbs being used as verbs in our dataset. Next, when Stanford assigns List Item (LS) and Symbol (SYM) to words in our data set, it is typically mis-annotating nouns, so we map these to a noun.

Finally, when we apply the Stanford tagger to function names, we append the letter \textit{I} to the beginning of the name. This is a known technique-- the Stanford+I technique, used to help Stanford tag identifiers that represent actions more accurately. It was used in previous studies applying part-of-speech tags to method identifier names \cite{Olney2016} to increase Stanford's accuracy. We also test to make certain that this did, in fact, improve Stanford's output and present these results later in Section~\ref{rqbdiscussion}. Other research has augmented input to Stanford or other NLP techniques improve analysis in the past \cite{Binkley:2011, abebe10}.

\subsection{Noun, Verb, and Prepositional phrases} \label{phrasalconcepts}
There are a few linguistic concepts that come up when we analyze part-of-speech tagger output. Specifically, we will be dealing with noun phrases, verb phrases, and prepositional phrases. We define these terms and give an example. A Noun Phrase (NP) is a sequence of noun modifiers, such as noun-adjuncts and adjectives, followed by a noun, and optionally followed by other modifiers or prepositional phrases \cite{Manning:1999}. The noun in a noun phrase is typically referred to as a \textit{head-noun}; the entity which is being modified/described by the words to its left \cite{Deissenbock:2005} (or, for programmers, sometimes surrounding it) in the phrase. A Verb Phrase (VP) is a verb followed by an NP and optionally a Prepositional Phrase (PP). A PP is a preposition plus an NP and can be part of a VP or NP.

Figure~\ref{fig:example_phrasal_concepts} presents an example NP, VP, and VP with PP for three method name identifiers. The phrase structure nodes are NP, VP, and PP, while the other nodes (i.e., N, NM, V, P) are part-of-speech annotations. The leaf nodes are the individual words split from within the identifier. Each word in the identifier is assigned a part-of-speech, which can then be used to derive the identifier’s phrase structure. We cannot build a phrase structure without part-of-speech information. One important thing to note about these phrases is how the words in the phrases work together. For example, in noun phrases, noun modifiers (e.g., other nouns or adjectives) work to modify (i.e., specify) the concept represented by the head-noun that is part of the same phrase. In Figure~\ref{fig:example_phrasal_concepts}, \textit{contentBorder} is a noun phrase where \textit{content} modifies our understanding of the noun \textit{border}. It tells us that we are talking about the content border as opposed to another type of border; a \textit{window border}, for example. When we make it into a verb phrase by adding draw to get \textit{drawContentBorder}; we add an action (i.e., draw) that will be applied to the particular type of border (i.e., the content border) represented by the identifier.

% \usepackage{booktabs}
\begin{table}[]
\centering
\caption{Total number per category of identifiers and unique grammar patterns across all systems}
\label{tab:systemstats}
\begin{tabular}{@{}lcc@{}}
\toprule
\multicolumn{1}{c}{Category} & \begin{tabular}[c]{@{}c@{}}Total Identifiers \\ Across All Systems\end{tabular} & \begin{tabular}[c]{@{}c@{}}\# of Unique Grammar\\ Patterns in Dataset\end{tabular} \\ \midrule
Decls & 920778 & 45 \\ \midrule
Classes & 37117 & 40 \\ \midrule
Functions & 428748 & 96 \\ \midrule
Parameters & 1197047 & 40 \\ \midrule
Attributes & 159562 & 53 \\ \midrule
\multicolumn{1}{c}{\textbf{Total}} & 2743252 & 277 \\ \bottomrule
\end{tabular}
\end{table}

\section{Methodology}
\label{methodology}
Identifiers come in many forms. The most common fall into one of the five following categories: class names, function names, parameter names, attribute names (i.e., data members), and declaration-statement names. A declaration-statement name is a name belonging to a function-local or global variable. We use this terminology as it is consistent with srcML's terminology \cite{collard:2016} for these variables and we used srcML to collect identifiers. Therefore, to study grammar patterns, we group a set of identifiers based on which of these five categories they fell into. The purpose of doing this is two-fold. 1) We can study grammar pattern frequencies based on their high-level semantic role (i.e., class names have a different role than function names). 2) We can consider the differences in accuracy for part-of-speech taggers when given identifiers from different categories.

\subsection{Setup for the dataset of human-annotated identifiers}
We created a gold set of grammar patterns for each of the five categories above by manually assigning (i.e., annotating) part-of-speech tags to each word within each identifier within each of the five categories above. We calculate the sample size by counting the total number of identifiers in each of the five categories (given in Table~\ref{tab:systemstats}) and calculating a sample based on a 95\% confidence level and a confidence interval of 6\%. We chose this confidence level and interval as a trade-off between time (i.e., annotating and validating is a manual effort) and accuracy. Using this confidence level and interval, we determine that each of our five categories needs to contain 267 samples (i.e., 267 was the largest number any of the sets required to be statistically representative, some required less-- we went with 267). This totals to 1335 identifiers in the entire set. This sample size is also similar to the number used in prior studies on part-of-speech tagging \cite{Olney2016,Gupta:2013}.

Initially, one author (annotator) was assigned to each category and was responsible for assigning grammar patterns for each of the 267 identifiers in the category. The annotators were given a split identifier (using Spiral \cite{HuckaSpiral}) along with the identifier's type and, if the identifier represented a function, the parameters/return types for that function. They were also allowed to look at the source code from which the identifier originated if needed. The annotators were asked to additionally identify and correct mistakes made by Spiral. 

We did not expand abbreviations. The reason for this is that abbreviation expansion techniques are not widely available (e.g., cannot be easily integrated into different languages or frameworks, cannot be readily trained, are not fully or publicly implemented) and still not very accurate \cite{newmanabbrev}. Therefore, a realistic worst-case scenario for developers and researchers is that no abbreviation-expansion technique available to use; their part-of-speech taggers must work in this worst-case scenario. We also tried not to split domain-term abbreviations (e.g., Spiral will make IPV4 into IPV 4; we corrected this back to IPV4). We did this because some taggers may recognize these domain terms. It is also the view of the authors that they should be recognized and appropriately tagged in their abbreviated (i.e., their most common) form. In the future, we plan to train a part-of-speech tagger using this dataset.
 
After completing their individual sets, the authors traded and reviewed one another's sets (i.e., performed cross-validation) twice. Thus, every identifier has been reviewed by two annotators. There was only one disagreement that could not be settled due to a particularly disfigured identifier; therefore, one identifier was randomly re-selected. This identifier is as follows: \textit{uint8x16\_t a\_p1\_q1\_2}; the annotators could not ascertain the meaning of the letters and numbers, making it difficult to tag. Once every identifier was assigned a grammar pattern manually and had been reviewed by at least two other authors, we ran each of our three part-of-speech taggers on the set of split identifiers; providing whatever information was required by the tagger (e.g., some taggers require full function signature, others only use the identifier name). We used srcML \cite{collard:2016} to obtain any additional information required by the taggers. The grammar pattern output from each tagger was used to generate frequency counts and compare to the manually-annotated grammar patterns to calculate accuracy. 

\subsection{Definition of Accuracy} \label{AccuracyDef}
Accuracy in this paper is synonymous with agreement; we compare the automatically generated annotations from the individual part-of-speech taggers with the manual annotations provided by humans. To be specific, we perform two different accuracy calculations for each tagger. One to determine the accuracy of each tagger on the individual part-of-speech tags found in Table \ref{tab:pertagagreement} and one to determine the accuracy of each tagger on full grammar patterns like those in Table \ref{tab:manualandtoolannotatedset}. To put this into an equation, we first define four sets. $H_{gp}$, the set of all human-annotated grammar patterns. $T_{gp}$, the set of all tool-annotated grammar patterns for a single part-of-speech tagger; there are three of these since we use three tools in this paper. $H_{word}$, the set of all human annotations for individual words in our set. Finally, $T_{word}$, the set of all tool annotations for individual words. Again, there are three of these since we use three part-of-speech tools in this paper. We then define grammar pattern level accuracy as the number of patterns which the human and tool sets agreed on (i.e., intersection) divided by the total number of grammar patterns annotated by humans. The equation follows:

\begin{equation}
    \begin{vmatrix}H_{gp}  \cap T_{gp} \end{vmatrix} \div  \begin{vmatrix}H_{gp}\end{vmatrix}
\end{equation}
We define word-level accuracy similarly. The number of words whose part-of-speech annotation was agreed upon by both humans and individual tools divided by the number of word-level human annotations. The equation follows:

\begin{equation}
    \begin{vmatrix}H_{word}  \cap T_{word} \end{vmatrix} \div  \begin{vmatrix}H_{word}\end{vmatrix} 
\end{equation}

To calculate $H_{gp} \cap T_{gp}$, we compare grammar pattern strings for individual identifiers from the human annotations with the corresponding tool annotations for the same identifier (i.e., using string matching) and take only exact string matches. To calculate $H_{word} \cap T_{word}$, we compare grammar pattern strings for individual identifiers the same way \textbf{except} we only look for exact string matches in the individual part-of-speech tags (i.e., for corresponding words) within the grammar pattern instead of requiring the full grammar pattern to match. For example, given two identifiers: \textit{get token string} and \textit{set factory handle}, which have a human annotated grammar pattern of \textit{V NM N}, if one tagger gives us the pattern \textit{NM NM N} then we would say that there is no grammar pattern intersection; the humans and tool gave different grammar patterns. However, there is an intersection here if we only look at individual part-of-speech tags. Both the tagger and humans annotated NM and N in the last two words of each identifier. Thus, these are considered matches and would be found in $H_{word} \cap T_{word}$. If a second tagger provided the grammar pattern \textit{V NM N}, then this would be found in $H_{gp} \cap T_{gp}$ and the individual annotation matches would be in $H_{word} \cap T_{word}$.

\begin{table}[]
\centering
\caption{Systems used to create dataset}
\label{tab:systemsizes}
\begin{tabular}{@{}lccc@{}}
\toprule
\multicolumn{1}{c}{\textbf{Name}} & \textbf{Size (kloc)} & \textbf{Age (years)} & \textbf{Language(s)} \\ \midrule
junit4 & 30 & 19 & Java \\ \midrule
mockito & 46 & 9+ & Java \\ \midrule
okhttp & 54 & 6 & Java \\ \midrule
antlr4 & 92 & 27 & Java/C/C++/C\# \\ \midrule
openFrameworks & 130 & 14 & C/C++ \\ \midrule
jenkins & 156 & 8 & Java \\ \midrule
irrlicht & 250 & 13 & C/C++ \\ \midrule
kdevelop & 260 & 19 & C/C++ \\ \midrule
ogre & 370 & 14 & C/C++ \\ \midrule
quantlib & 370 & 19 & C/C++ \\ \midrule
coreNLP & 582 & 6 & Java \\ \midrule
swift & 601 & 5 & C++/C \\ \midrule
calligra & 660 & 19 & C/C++ \\ \midrule
gimp & 777 & 23 & C/C++ \\ \midrule
telegram & 912 & 6 & Java/C/C++ \\ \midrule
opencv & 1000 & 19 & C/C++ \\ \midrule
elasticsearch & 1300 & 9 & Java \\ \midrule
bullet3 & 1300 & 10+ & C/C++/C\# \\ \midrule
blender & 1600 & 21 & C/C++ \\ \midrule
grpc & 1800 & 5 & C++/C/C\# \\ \bottomrule
\end{tabular}
\end{table}

\subsection{Data Collection}
\label{datacollection}
We collected our identifier set from twenty open-source systems. We chose these systems to vary in terms of size and programming language while also being mature and having their own development communities. We did this to make sure that the identifiers in these systems have been seen by multiple contributors and that the identifiers we collected are not biased toward a specific programming language. There are two reasons for choosing identifiers from multiple languages. 1) We want to know what patterns cross-cut between languages, such that most Java/C/C++ developers are familiar with and leverage these patterns. Focusing on just one language might mean the patterns we find are not common to developers outside of the chosen language. 2) Many systems are written in more than one language, and it is important to understand how well part-of-speech tagging technologies will work on these systems. Thus, running our study systems written in different programming languages helps us study part-of-speech tagger results in an environment leveraging multiple programming languages. This does not mean that none of our patterns are biased to one language or system, but that the most frequent patterns are less likely to be; we confirm cross-cutting patterns in Section \ref{evaluation}.

We provide the list of systems and their characteristics in Table~\ref{tab:systemsizes}. The systems we picked were 615 KLOC on average with a median of 476 KLOC, a min of 30 KLOC, and a max of 1,800 KLOC. Further, most of these systems have been in active development for the past ten years or more and all of them for five years or more. The younger systems in our set are popular, modern programs. For example, Swift is a well-known programming language supported by Apple, Telegram is a popular messaging app, and Jenkins is a popular development automation server. Because we are trying to measure the accuracy of part-of-speech techniques and understand common grammar patterns, our goal is not necessarily to study only high-quality identifier names, but to study names that are closely representative of the average name for open-source systems. Additionally, we remove identifiers that appear in test files, in part because they sometimes have specialized naming conventions (e.g., include the word `test', `assert', `should', etc). We exclude test-related identifiers by ignoring annotated test files and directories; any directory, file, class, or function containing the word \textit{test}. While it is possible that identifiers in test code have similar grammar patterns to identifiers outside of test code, it is also possible that they do not. We did not want to risk introducing divergent grammar patterns. We think it would be appropriate to study test identifier grammar patterns separately to confirm their similarity, or dissimilarity, to other identifiers.

To collect the 1,335 identifiers, we scanned each of our 20 systems using srcML \cite{collard:2016} and collected both identifier names/types and the category that they fell into (e.g., class, function). Then, for each category, we randomly selected one identifier from each system using a round-robin algorithm (i.e., we picked a random identifier from system 1, then randomly selected an identifier from system 2, etc. until we hit 267). This ensured that we got either 13 or 14 identifiers from each system (267/20 = 13.35) per category and mitigates the threat of differing system size.

We provide all data that was part of this study on our webpage \footnote{https://scanl.org/} for replication, extension, and for use in further evaluation of part-of-speech taggers for source code.

\begin{table}[]
\centering
\caption{Top 5 patterns in dataset along with frequency of each pattern and \% of the set represented by that pattern}
\label{tab:manualandtoolannotatedset}
\resizebox{!}{.28\paperheight}{%
\begin{tabular}{@{}llllllll@{}}
\toprule
\multicolumn{8}{c}{\textbf{Attribute Names}} \\ \midrule
\multicolumn{2}{c|}{Humans} & \multicolumn{2}{c|}{Posse} & \multicolumn{2}{c|}{Swum} & \multicolumn{2}{c|}{Stanford} \\ \midrule
NM N & \multicolumn{1}{l|}{78 (29.2\%)} & N N & \multicolumn{1}{l|}{82 (30.7\%)} & NM N & \multicolumn{1}{l|}{122 (45.7\%)} & N N & 59 (22.1\%) \\ \midrule
NM NM N & \multicolumn{1}{l|}{34 (12.7\%)} & N N N & \multicolumn{1}{l|}{35 (13.1\%)} & NM NM N & \multicolumn{1}{l|}{63 (23.6\%)} & N N N & 26 (9.7\%) \\ \midrule
NM NPL & \multicolumn{1}{l|}{26 (9.7\%)} & NM N & \multicolumn{1}{l|}{31 (11.6\%)} & NM NM NM N & \multicolumn{1}{l|}{32 (12\%)} & NM N & 18 (6.7\%) \\ \midrule
N & \multicolumn{1}{l|}{16 (6\%)} & N & \multicolumn{1}{l|}{26 (9.7\%)} & N & \multicolumn{1}{l|}{27 (10.1\%)} & V N & 16 (6\%) \\ \midrule
NM NM NM N & \multicolumn{1}{l|}{11 (4.1\%)} & NM N N & \multicolumn{1}{l|}{16 (6\%)} & NM NM NM NM N & \multicolumn{1}{l|}{11 (4.1\%)} & N NPL & 16 (6\%) \\ \midrule
\multicolumn{8}{c}{\textbf{Declaration Names}} \\ \midrule
NM N & \multicolumn{1}{l|}{116 (43.4\%)} & N N & \multicolumn{1}{l|}{112 (41.9\%)} & NM N & \multicolumn{1}{l|}{164 (61.4\%)} & N N & 60 (22.5\%) \\ \midrule
NM NM N & \multicolumn{1}{l|}{43 (16.1\%)} & NM N & \multicolumn{1}{l|}{41 (15.4\%)} & NM NM N & \multicolumn{1}{l|}{69 (25.8\%)} & NM N & 33 (12.4\%) \\ \midrule
NM NPL & \multicolumn{1}{l|}{30 (11.2\%)} & N N N & \multicolumn{1}{l|}{30 (11.2\%)} & NM NM NM N & \multicolumn{1}{l|}{17 (6.4\%)} & V N & 24 (9\%) \\ \midrule
NM NM NPL & \multicolumn{1}{l|}{8 (3\%)} & NM N N & \multicolumn{1}{l|}{19 (7.1\%)} & N & \multicolumn{1}{l|}{6 (2.2\%)} & N N N & 21 (7.9\%) \\ \midrule
NM NM NM N & \multicolumn{1}{l|}{6 (2.2\%)} & N & \multicolumn{1}{l|}{6 (2.2\%)} & DT NM N & \multicolumn{1}{l|}{4 (1.5\%)} & N NPL & 15 (5.6\%) \\ \midrule
\multicolumn{8}{c}{\textbf{Parameter Names}} \\ \midrule
NM N & \multicolumn{1}{l|}{122 (45.7\%)} & N N & \multicolumn{1}{l|}{96 (36\%)} & NM N & \multicolumn{1}{l|}{155 (58.1\%)} & N N & 62 (23.2\%) \\ \midrule
NM NM N & \multicolumn{1}{l|}{36 (13.5\%)} & NM N & \multicolumn{1}{l|}{38 (14.2\%)} & NM NM N & \multicolumn{1}{l|}{63 (23.6\%)} & V N & 32 (12\%) \\ \midrule
N & \multicolumn{1}{l|}{21 (7.9\%)} & N N N & \multicolumn{1}{l|}{28 (10.5\%)} & N & \multicolumn{1}{l|}{30 (11.2\%)} & NM N & 31 (11.6\%) \\ \midrule
NM NPL & \multicolumn{1}{l|}{20 (7.5\%)} & N & \multicolumn{1}{l|}{23 (8.6\%)} & NM NM NM N & \multicolumn{1}{l|}{11 (4.1\%)} & N N N & 20 (7.5\%) \\ \midrule
NM NM NPL & \multicolumn{1}{l|}{12 (4.5\%)} & NM N N & \multicolumn{1}{l|}{16 (6\%)} & DT N & \multicolumn{1}{l|}{4 (1.5\%)} & N & 15 (5.6\%) \\ \midrule
\multicolumn{8}{c}{\textbf{Function Names}} \\ \midrule
V NM N & \multicolumn{1}{l|}{46 (17.2\%)} & V N N & \multicolumn{1}{l|}{49 (18.4\%)} & V NM N & \multicolumn{1}{l|}{66 (24.7\%)} & V N N & 42 (15.7\%) \\ \midrule
V N & \multicolumn{1}{l|}{26 (9.7\%)} & V N & \multicolumn{1}{l|}{40 (15\%)} & V N & \multicolumn{1}{l|}{41 (15.4\%)} & V N & 33 (12.4\%) \\ \midrule
V NM NM N & \multicolumn{1}{l|}{17 (6.4\%)} & V NM N & \multicolumn{1}{l|}{20 (7.5\%)} & V NM NM N & \multicolumn{1}{l|}{33 (12.4\%)} & V N N N & 19 (7.1\%) \\ \midrule
NM N & \multicolumn{1}{l|}{15 (5.6\%)} & V N N N & \multicolumn{1}{l|}{15 (5.6\%)} & V NM NM NM N & \multicolumn{1}{l|}{14 (5.2\%)} & N N N & 8 (3\%) \\ \midrule
V NM NPL & \multicolumn{1}{l|}{10 (3.7\%)} & V NM N N & \multicolumn{1}{l|}{10 (3.7\%)} & NM N & \multicolumn{1}{l|}{13 (4.9\%)} & NM N & 8 (3\%) \\ \midrule
\multicolumn{8}{c}{\textbf{Class Names}} \\ \midrule
NM N & \multicolumn{1}{l|}{81 (30.3\%)} & N N & \multicolumn{1}{l|}{76 (28.5\%)} & NM NM N & \multicolumn{1}{l|}{98 (36.7\%)} & N N & 71 (26.6\%) \\ \midrule
NM NM N & \multicolumn{1}{l|}{72 (27\%)} & N N N & \multicolumn{1}{l|}{59 (22.1\%)} & NM N & \multicolumn{1}{l|}{94 (35.2\%)} & N N N & 69 (25.8\%) \\ \midrule
NM NM NM N & \multicolumn{1}{l|}{16 (6\%)} & NM N N & \multicolumn{1}{l|}{23 (8.6\%)} & NM NM NM N & \multicolumn{1}{l|}{39 (14.6\%)} & N N N N & 23 (8.6\%) \\ \midrule
N & \multicolumn{1}{l|}{14 (5.2\%)} & NM N & \multicolumn{1}{l|}{19 (7.1\%)} & N & \multicolumn{1}{l|}{16 (6\%)} & N & 13 (4.9\%) \\ \midrule
PRE NM N & \multicolumn{1}{l|}{10 (3.7\%)} & N N N N & \multicolumn{1}{l|}{15 (5.6\%)} & NM NM NM NM N & \multicolumn{1}{l|}{8 (3\%)} & NM N & 9 (3.4\%) \\ \bottomrule
\end{tabular}
}
\end{table}

%Generally, Posse and Stanford generated fewer instances of this pattern than Swum or the human annotators, so while there is some agreement, it is far from universal. For Swum and the human annotators, \textit{NM N} was the most common pattern except for in function names and parameter names.

\section{Evaluation} \label{evaluation}
Our evaluation aims to 1) establish and explore the common and diverse grammar pattern structures found in identifiers. 2) Use these structures to investigate the accuracy, strengths, and weaknesses of approaches which generate grammar patterns with an eye toward establishing and improving their current ability. And 3) leverage the grammar patterns we discover to discuss the ways in which words, as part of a larger identifier, work together to convey information to the developer. We address these concerns in the research questions that follow. For the discussion of RQs below, the \textit{+} symbol means ``one or more" and the \textit{*} symbol means ``zero or more" of the annotation to the left of the symbol; similar to how they are used in regular expressions.

\begin{figure}[t]%[H]
 	\centering
 	\includegraphics[width=1\linewidth]{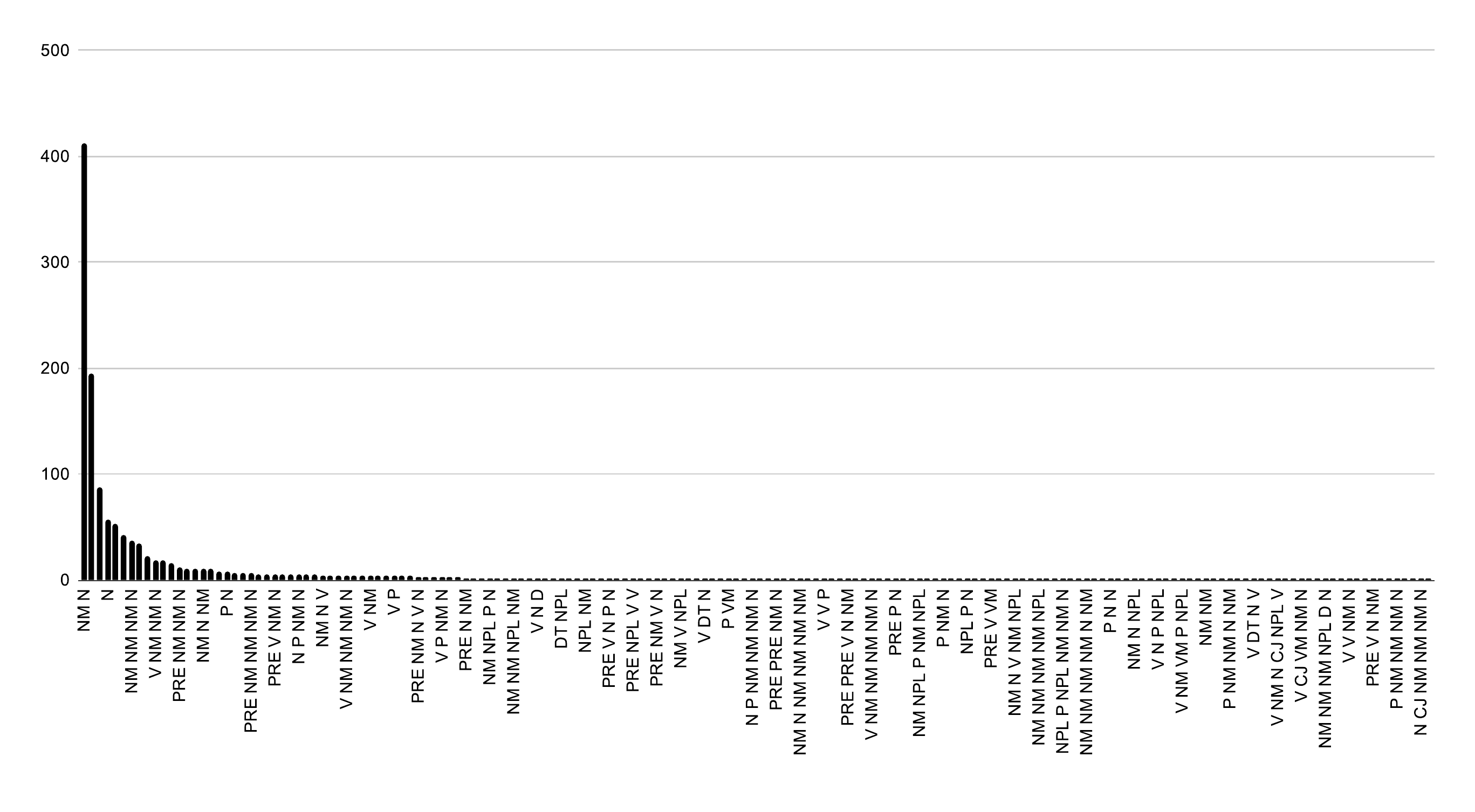}
 	\caption{Distribution of unique grammar pattern frequency over entire dataset -- not all unique patterns are shown due to space.}
 	\label{Figure:gpf}
\end{figure}

\begin{figure}[t]%[H]
 	\centering
 	\includegraphics[width=1\linewidth]{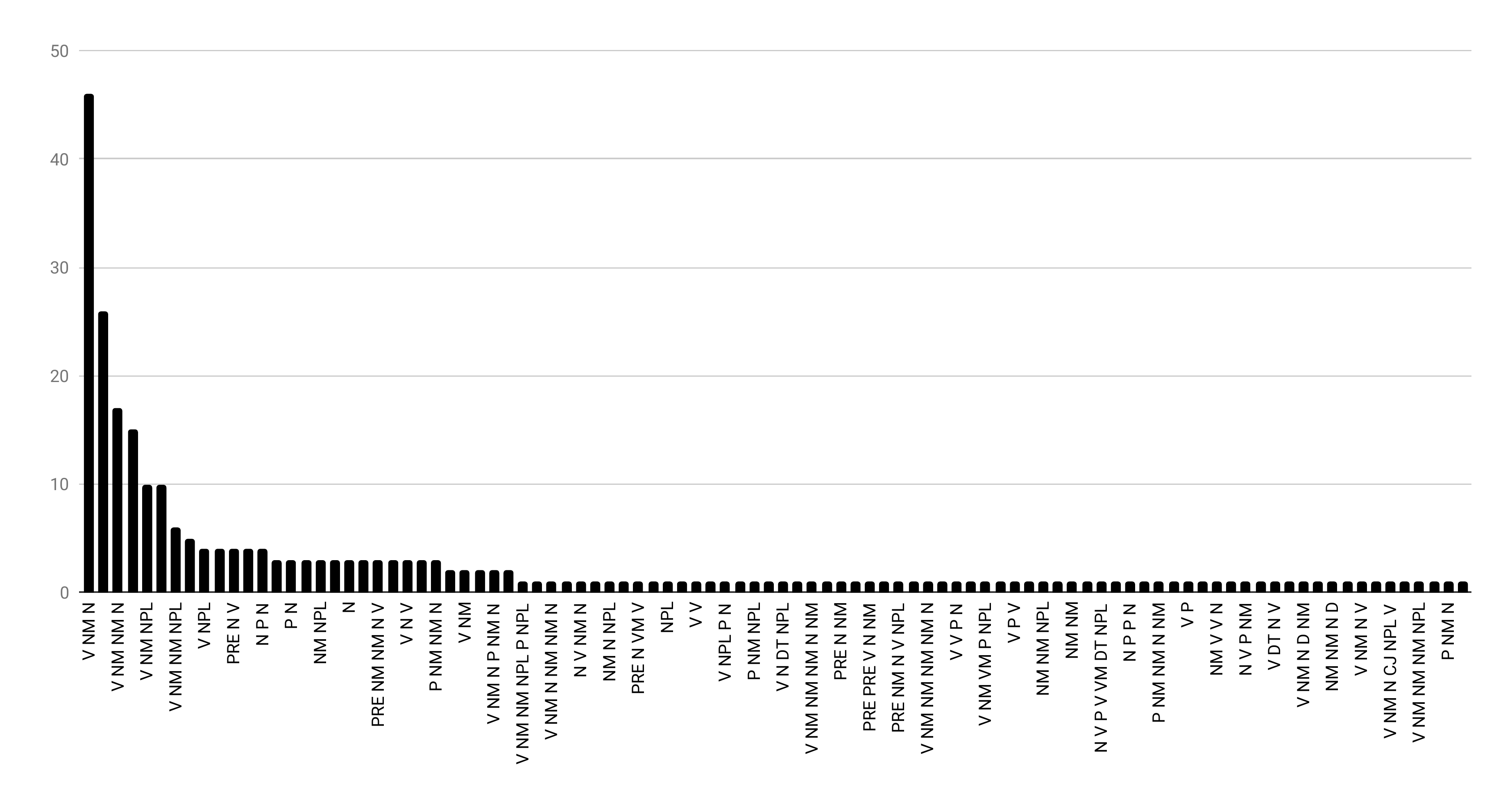}
 	\caption{Distribution of unique method grammar pattern frequency -- not all unique patterns are shown due to space.}
 	\label{Figure:gpf_methods}
\end{figure}
\subsection{RQ1: \RQA} \label{rqadiscussion}
Table \ref{tab:manualandtoolannotatedset} contains data from the human-annotated set separated into categories based on the location of the identifier in code. The table shows the top five grammar patterns for each category and each of the three taggers we used to generate the grammar patterns. In addition, Figure \ref{Figure:gpf} shows the distribution of unique grammar patterns across all identifiers. It shows that a minority of the total grammar patterns found in the set are repeated frequently, while the majority occur only once. Due to the fact that functions had the highest number of unique grammar patterns (Table \ref{tab:systemstats}), we also show the distribution of unique function grammar patterns in Figure \ref{Figure:gpf_methods}. The distribution is largely similar to the prior figure with all unique grammar patterns, but the most common function grammar patterns are different than the general set due to the semantics being conveyed by function names (e.g., use of verbs to convey actions) versus other identifiers. There are three grammar patterns from which most frequent grammar patterns we will discuss are derived. These are shown and described in Table \ref{tab:commontable}. Specifically, they are the verb phrase, noun phrase, and plural noun phrase patterns. We now discuss the most common grammar patterns found in our data set.
%\emily{Looks great! Could even stack them as subfigures to really make the differences pop -- make their height shorter, but not change the width. Or, time permitting, could try a single stacked bar chart with ordering from the first figure that shows the whole.}

\begin{table}[]
\centering
\caption{Grammar patterns from which other grammar patterns are frequently derived.}
\label{tab:commontable}
\resizebox{\textwidth}{!}{%
\begin{tabular}{@{}cl@{}}
\toprule
\textbf{Grammar Pattern} &
  \multicolumn{1}{c}{\textbf{Pattern Semantics}} \\ \midrule
\textbf{NM+ N} &
  \begin{tabular}[c]{@{}l@{}}Noun phrase pattern: One or more noun modifiers (adjectives or noun-adjuncts) that modify a single head-noun. \\ Noun modifiers are used to modify developers' understanding of the entity referred to by the head-noun. Because \\ identifier names are typically associated with a single entity (e.g., a list entity, a character entity, a string entity), \\ the head-noun typically refers to this entity. The noun modifiers, which are typically to the left the head-noun, are \\ used to specify characteristics, identify a context, or otherwise help the developer gain a stronger, more specific \\ understanding of what this entity embodies.\end{tabular} \\ \midrule
\textbf{V NM+ N} &
  \begin{tabular}[c]{@{}l@{}}Verb phrase pattern: A verb is followed by a noun phrase. Verb phrases in source code combine the action of a\\ verb with the descriptiveness of noun phrases; the verb specifies the action and the noun phrase contains both\\ the entity (i.e., the head-noun) which will be acted upon as well as noun modifiers which specify characteristics,\\ identify a context, or otherwise help the developer gain a stronger, more specific understanding of what this\\ entity embodies.\end{tabular} \\ \midrule
\textbf{NM+ NPL} &
  \begin{tabular}[c]{@{}l@{}}Plural noun phrase pattern: Similar to a regular noun phrase category (NM+ N), but the head-noun is plural instead\\ of singular. This is sometimes purposeful; used to refer to arrays, lists, and other collection types or used\\ to refer to the multiplicity of heterogeneous data groupings (i.e., classes/objects).\end{tabular} \\ \bottomrule
\end{tabular}%
}
\end{table}

\textbf{Grammar Pattern \textit{PRE* NM+ N}:} Looking at Table~\ref{tab:manualandtoolannotatedset}, the \textit{NM N} instance of this pattern appears in the top five most frequent in each category, and it is the most ubiquitous pattern in our dataset. Noun modifiers are used to modify developers' understanding of the entity referred to by the head-noun (Table~\ref{tab:commontable}). \textbf{Examples}: \textit{factory class}, \textit{auth result}, and \textit{previous caption}. These identifiers embody the noun phrase concept. The specific entities that they reference are \textit{class}, \textit{result}, and \textit{caption} respectively. While the noun modifiers (factory, auth, and previous) specify some characteristics of the entity they comes before. The \textit{class} is not just any kind of class; it is a \textit{factory} class, the \textit{result} is specifically the consequence of some form of \textit{authentication}, the \textit{caption} being referred to is the \textit{previous} in relation to some other caption. Adding more noun modifiers emphasizes this effect. \textbf{Examples:} \textit{max buffer size} uses \textit{max} and \textit{buffer} to describe characteristics of the \textit{size}, which is the head-noun. The same applies to \textit{previous initial value} and \textit{network security policy}.

This pattern is found in all categories, but it is somewhat out of place in the Function Name category, since function names tend to contain a verb. We manually investigated these instances and found that many of them are functions with an implied verb. \textbf{Examples:} \textit{deep Stub}. It contains no verb, but its behavior is to return (i.e., \textit{get}) a \textit{deep stub} object. Another example is the \textit{lower Boundary Factor} function which returns the \textit{lower boundary factor} based on a parameter. Many of these functions are getters or setters where the \textit{get} or \textit{set} is not in the name of the function.

The PRE* at the front of this pattern is optional, but appears frequently in the Class Name category. \textit{PRE} represents preambles, as defined in Section \ref{grammarpatterndef}. They are important to detect because if we cannot identify preambles automatically, tools may assume that the preamble somehow augments developer understanding of the head noun-- which it does not. Therefore, preamble detection can help tools avoid making mistakes in interpreting words in an identifier. We discuss more about preamble patterns in Section \ref{rqbdiscussion}.

The ubiquity of noun-phrase patterns suggests that part-of-speech taggers should predict \textit{NM} on unknown, non-numeric tokens which are not the last token in the identifier-- for the last token, \textit{N} would be a better prediction. 

%One final interesting note is that the noun modifiers in extended noun phrases can be combined (e.g., abbreviated) into one noun modifier and potentially not lose their effectiveness with respect to helping developers comprehend. For example, if we abbreviate the noun modifiers in \textit{max buffer size}, we might get \textit{mb size} which would have a grammar pattern of \textit{NM N}; the meaning has not changed. This is routine to developers who frequently abbreviate terms without loss of meaning as long as other developers recognize the abbreviation. 
\textbf{Grammar Pattern \textit{V NM+ N}:} This pattern and its extensions are most common in functions. This is a verb phrase pattern, where a verb is followed by a noun phrase (Table~\ref{tab:commontable}). \textbf{Examples}: \textit{check gl support}, \textit{resize nearest neighbor}, and \textit{get max shingle diff}. In \textit{check gl Support}, the noun phrase specifies what we are interested in (openGL support), \textit{check} tells us that we are checking a condition-- specifically that \textit{gl support} is available. The same applies for the other two identifiers; specifying which neighbor to \textit{resize} in \textit{resize nearest neighbor} and the nature of the \textit{diff[erence]} to return (i.e., \textit{get}) in \textit{get max shingle diff}.

\begin{table}[]
\centering
\caption{Frequency at which identifiers with a verb in their grammar pattern also have a boolean type and frequency at which identifiers with a boolean type have a verb in their grammar pattern}
\label{tab:booleanverbtable}
\resizebox{\textwidth}{!}{%
\begin{tabular}{@{}lccccc@{}}
\toprule
 & \textbf{\begin{tabular}[c]{@{}c@{}}Identifier\\ Contains Verb\end{tabular}} & \textbf{\begin{tabular}[c]{@{}c@{}}Identifier Has\\ Boolean Type\end{tabular}} & \textbf{\begin{tabular}[c]{@{}c@{}}Identifier Has Boolean \\ Type \& Contains Verb\end{tabular}} & \textbf{\begin{tabular}[c]{@{}c@{}}\% of Boolean Identifiers\\ Containing a Verb\end{tabular}} & \textbf{\begin{tabular}[c]{@{}c@{}}\% of Verb Identifiers\\ With a Boolean Type\end{tabular}} \\ \midrule
\textbf{Parameters} & 24 & 28 & 22 & 92\% & 79\% \\ \midrule
\textbf{Declarations} & 21 & 21 & 18 & 86\% & 86\% \\ \midrule
\textbf{Attributes} & 23 & 24 & 18 & 78\% & 75\% \\ \bottomrule
\end{tabular}
}
\end{table}
One characteristic we observed about this and other verb-based patterns is that, when it appears outside of function names, it tends to be for an identifier with a boolean type or a type which can be treated as boolean (e.g., integer). \textbf{Examples}: \textit{add bias to embedding}, \textit{is first frame}, and \textit{will return last parameter}. This is because boolean variables act like predicates; asking a question whose answer is true or false (e.g., add bias to [the] embedding?, is [this] the first frame?, will [this] return [the] last parameter?). Other researchers have made this observation \cite{Binkley:2011, Gupta:2013}, but only one reports quantity \cite{butler:2015}. In Table~\ref{tab:booleanverbtable}, we show two things: 1) the percentage of all identifiers in the dataset with a at least one verb in their grammar pattern and a type which can be interpreted as boolean. 2) the percentage of all identifiers in the dataset with a boolean type that also have a at least one verb in their grammar pattern. 

The observation is supported, especially amongst parameter and declaration-statement identifiers where 92\% (22/24) and 86\% (18/21) respectively of all identifiers with grammar pattern containing a verb also have a boolean type. Likewise, of all parameters and declaration-statement identifiers with a boolean type, 79\% (22/28) and 86\% (18/21) respectively contain a verb in their grammar pattern. Given that a part-of-speech tagger can be made aware of a given identifier's type, this trend should be useful in helping properly annotate boolean variables as well as suggesting higher-quality names for boolean variables.

\begin{table}[]
\centering
\caption{Frequency at which identifiers ending with a plural also have a collection type and frequency at which identifiers with a collection type end with a plural}
\label{tab:pluralidentifiertable}
\resizebox{\textwidth}{!}{%
\begin{tabular}{@{}lccccc@{}}
\toprule
 & \textbf{\begin{tabular}[c]{@{}c@{}}Identifier has\\ Collection Type\end{tabular}} & \textbf{\begin{tabular}[c]{@{}c@{}}Identifier Ends\\ with Plural\end{tabular}} & \textbf{\begin{tabular}[c]{@{}c@{}}Identifier Ends with Plural\\ \& Has Collection Type\end{tabular}} & \textbf{\begin{tabular}[c]{@{}c@{}}\% of Identifiers w/Collection\\ Type \& End With Plural\end{tabular}} & \textbf{\begin{tabular}[c]{@{}c@{}}\% of Identifiers w/Plural \\ Ending \& Collection Type\end{tabular}} \\ \midrule
\textbf{Parameters} & 49 & 42 & 21 & 43\% & 50\% \\ \midrule
\textbf{Decls} & 56 & 49 & 24 & 43\% & 49\% \\ \midrule
\textbf{Attributes} & 43 & 61 & 31 & 72\% & 51\% \\ \midrule
\textbf{Functions} & 21 & 44 & 10 & 48\% & 23\% \\ \bottomrule
\end{tabular}
}
\end{table}
\textbf{Grammar Pattern \textit{V* NM+ NPL}:} This pattern has two configurations. It is either a plural noun phrase pattern or a plural verb phrase pattern (Table~\ref{tab:commontable}). Assuming that the use of a plural must be significant somehow, since some sources advise using plural identifiers for collections and certain types of classes \cite{Ambler:1999}, we analyzed our dataset to see if identifiers which have a plural head-noun were more likely to have a type which indicated a collection (e.g., list, array) type. To do this, we examine the type name for each identifier and record if it contains the words: \textit{list, map, dictionary, collection, array, vector, data, or set}. We additionally record if the type name is plural (e.g., ending in -s) or if the identifier has square brackets (i.e., []) next to it. We then manually check these to see if they were really collection types. The results of this investigation are in Table \ref{tab:pluralidentifiertable}. This table shows two perspectives on the data: 1) how many identifiers with a collection type also have a plural head-noun. 2) how many identifiers that have a plural head-noun also have a collection type.

We found that of all identifiers with a collection type, 43\%, 43\%, 72\%, and 48\% of Parameter, Declaration-statement, Attribute, and Function identifiers, respectively, are also plural. Additionally, of all identifiers that are plural, 50\%, 49\%, 51\%, and 23\% of Parameter, Declaration-statement, Attribute, and Function identifiers, respectively, have a collection type. Similar to booleans, we find that there is a trend-- particularly for attribute identifiers with a collection type, which had the highest likelihood of also being plural. While this is not always the majority case, it does suggest an interesting direction for future research into the use of plural names to convey the use of collections in different types of identifiers. \textbf{Examples}: \textit{mkt factors}, \textit{num cols}, and \textit{child categories}. The \textit{mkt factors} identifier is a plural noun phrase that represents a collection entity (e.g., a list or array) of market factors. The \textit{num cols} identifier represents the number of columns for some entity (e.g., a matrix), and the \textit{child categories} identifier represents a set of categories with a parent-child relationship to a super-category.

The weakest correlation between use of plural noun phrases and collection type is found in the Function Name category as part of a plural verb phrase. Instead of representing a collection, Function identifiers following a plural verb phrase pattern often allude to the multiplicity of the data being operated on, and not necessarily returned. A few examples from our data set are: \textit{Object getRawArguments(...)}, \textit{void validateClassRules(...)}, and \textit{Object getActualValues(...)}. In all three cases, while a collection is not explicitly returned, the functions refer to an entity which represents multiple heterogeneous data (i.e., an object) that is then returned, part of the calling object, or incoming data as a parameter. This behavior is not unique to functions, but more frequent in function identifiers versus other identifiers. Therefore, we should be cautious when making assumptions.

\textbf{Grammar Pattern \textit{V N}:} This pattern represents an action applied to or with the support of an entity represented by the head-noun. This pattern commonly represents function names or boolean identifiers, similar to the \textit{V NM N} pattern from which it differs only due to the lack of a noun-adjunct. \textbf{Example:} \textit{read subframe} has the grammar pattern \textit{V N} and names a function which reads a subframe object.

%This pattern, like \textit{V NM N}, is also popular with boolean variables for the same reason. Its relationship with \textit{V NM N} is simply that the developer did not feel that an adjective was needed to help describe the noun; the noun was descriptive enough on its own.

%\textbf{Grammar Pattern \textit{V}:} Like the other patterns derived from verbs above, this one is typically used with boolean variables and functions. One interesting thing about this pattern is that there there is no noun for the verb to act on. Typically, when this happens, it is because the noun is contained within its arguments or is the calling object itself (i.e., \textit{this}). Some examples are the identifiers \textit{visit} and \textit{intersect}, where the \textit{visit} function can be applied to a tree data structure and the \textit{intersect} function can be applied to two or more sets to find their intersection. Notice that the full grammar pattern for these actually depends on the number of arguments they take. For example, since \textit{intersect} works two or more sets, the size of the grammar pattern formed by its name and all of the sets it will operate on is linear with the number of sets.

\textbf{Grammar Pattern \textit{N}:} A single noun, trivially a head-noun, represents a singular entity and could be considered a basic case of the noun phrase pattern (i.e., with no NMs). \textbf{Example}: \textit{client} and \textit{usage}. Poorly split, or purposefully not split, abbreviations are automatically tagged as a single noun. This phenomenon is not necessarily incorrect as it could be considered a collapsed noun-phrase pattern (e.g., \textit{max buffers} abbreviated as \textit{mb} which then gets annotated as a noun). Developers familiar with the abbreviations may even prefer this form and certain, extremely well known, abbreviations may even be treated as singular nouns during comprehension. For example, IPV4, MP3, HTTP, HTML are common abbreviations which are more common to see/read than their expansions (e.g., MP3 is rarely expanded).

%One interesting thing we noticed in the dataset is that sometimes developers use a noun modifier (e.g., an adjective) as a noun, which would not make sense in English. An example from our dataset is the identifier \textit{actual}, which is not a noun, but instead an adjective being used as a noun. However, this word makes sense to programmers. The reason for this is that this is a noun phrase, but the noun is not written in the identifier name. That is, \textit{actual} is an abbreviation of sorts; an adjective missing its head-noun. With a head-noun, we would have identifiers such as: \textit{actual parameter}, \textit{actual value} or \textit{actual result}.

\textbf{\textit{{Summary for RQ1}}}: Table~\ref{tab:manualandtoolannotatedset} contains the most frequent grammar patterns in the human-annotated set. We identified five patterns by looking at how frequently they occurred in our human-annotated dataset. By far, the most ubiquitous pattern we found was the noun phrase (\textit{NM+ N}) pattern as it appears in every category. We also found that verb phrase (\textit{V NM+ N)} patterns are most common in function names but also appear in other types of identifiers; specifically those with a boolean type, and that plural noun phrases (\textit{NM+ NPL}) have a somewhat heightened chance of representing collection identifiers. Results indicate that 1) due to the ubiquity of noun phrase patterns, part-of-speech taggers should predict \textit{NM} on unknown, non-numeric tokens that are not the right-most token in the identifier; \textit{N} is a better prediction for the right-most token, as it is likely the head-noun. 2) our data supports the observed, heightened appearance of verbs in boolean variables from prior work. 3) while there is a link between identifier names containing a plural and collection data types, the plural is sometimes used to reference multiplicity of related, heterogeneous data (i.e., class member data), particularly for function identifiers. This presents an opportunity to support developers in how, and when, to use plurals. The grammar patterns we generated highlight how the name of an identifier is influenced by the semantics of the language and gives us a glimpse into how developers use words in an identifier to comprehend their code.

\begin{table}[]
\centering
\caption{Frequency of per-tag agreement between human annotations and tool annotations}
\label{tab:pertagagreement}
\resizebox{\textwidth}{!}{%
\begin{tabular}{@{}lccccccc@{}}
\toprule
\textbf{\begin{tabular}[c]{@{}l@{}}Part of \\ Speech\end{tabular}} & \textbf{\begin{tabular}[c]{@{}c@{}}Human\\ Annotations\end{tabular}} & \textbf{Posse} & \textbf{\begin{tabular}[c]{@{}c@{}}\% Agreement\\ w/ Humans\end{tabular}} & \textbf{Swum} & \textbf{\begin{tabular}[c]{@{}c@{}}\% Agreement\\ w/ Humans\end{tabular}} & \textbf{Stanford} & \textbf{\begin{tabular}[c]{@{}c@{}}\% Agreement\\ w/ Humans\end{tabular}} \\ \midrule
NM & 1604 & 373 & 23.25\% & 1508 & \textbf{94.01\%} & 252 & 15.71\% \\ \midrule
N & 1141 & 1025 & 89.83\% & 976 & 85.54\% & 1064 & \textbf{93.25\%} \\ \midrule
V & 305 & 205 & 67.21\% & 171 & 56.07\% & 233 & \textbf{76.39\%} \\ \midrule
NPL & 238 & 0 & 0.00\% & 0 & 0.00\% & 171 & \textbf{71.85\%} \\ \midrule
PRE & 105 & 0 & 0.00\% & 2 & \textbf{1.90\%} & 0 & 0.00\% \\ \midrule
P & 94 & 59 & 62.77\% & 28 & 29.79\% & 85 & \textbf{90.43\%} \\ \midrule
D & 27 & 0 & 0.00\% & 5 & 18.52\% & 27 & \textbf{100.00\%} \\ \midrule
DT & 15 & 6 & 40.00\% & 13 & \textbf{86.67\%} & 9 & 60.00\% \\ \midrule
VM & 13 & 0 & 0.00\% & 0 & 0.00\% & 9 & \textbf{69.23\%} \\ \midrule
CJ & 8 & 0 & 0.00\% & 0 & 0.00\% & 4 & \textbf{50.00\%} \\ \midrule
\multicolumn{1}{l}{\textit{\textbf{Total words:}}} & \textit{\textbf{3550}} & \textit{\textbf{1668}} & \textit{\textbf{}} & \textit{\textbf{2703}} & \textit{\textbf{}} & \textit{\textbf{1854}} & \multicolumn{1}{l}{\textit{\textbf{}}} \\ \bottomrule
\end{tabular}
}
\end{table}

\subsection{RQ2: \RQB} \label{rqbdiscussion}
We compare the output of the three part-of-speech taggers in this study with the manually-annotated grammar patterns in order to calculate the accuracy of each tagger at both the level of grammar patterns and the level of individual part-of-speech tags. Please refer to Section \ref{AccuracyDef} for an explanation of how we calculate accuracy. Starting with Table~\ref{tab:pertagagreement}, which contains our per-tag (i.e., word-level) accuracy analysis, we observe that Swum had the highest agreement with the human annotations with respect to noun modifiers, determiners, and preambles. Stanford had the highest agreement with respect to everything else. Posse never outperformed both the other two in a single category, but did perform better than Stanford at annotating noun modifiers and better than Swum at annotating nouns, prepositions, and verbs. The numbers here indicate that Stanford has the best all-around performance when we are looking at accuracy on individual part-of-speech annotations, as it is able to detect a broader range of them more accurately than either Swum or Posse. However, while Stanford had the highest accuracy on the largest number of part-of-speech tag types, Swum tagged the highest number of raw words correctly with 2,703 correct annotations versus Stanford's 1,854. The results indicate areas of strength for each tagger; their combined output may increase their overall accuracy.

\begin{table}[]
\centering
\caption{Percentage of tool-annotated grammar patterns which fully match human-annotated grammar patterns}
\label{tab:accuracytable}
\resizebox{\textwidth}{!}{%
\begin{tabular}{@{}lllllll@{}}
\toprule
\textbf{Category} & \multicolumn{1}{c}{\textbf{Posse}} & \multicolumn{1}{c}{\textbf{Swum}} & \multicolumn{1}{c}{\textbf{Stanford (V)}} & \textbf{Stanford (NM)} & \textbf{Stanford+I (V)} & \textbf{Stanford+I (NM)} \\ \midrule
\textbf{Parameters} & 58 (21.7\%) & 181 (67.8\%) & 63 (23.6\%) & 71 (26.6\%) & N/A & N/A \\ \midrule
\textbf{Declarations} & 47 (17.6\%) & 163 (61\%) & 51 (19.1\%) & 54 (20.2\%) & N/A & N/A \\ \midrule
\textbf{Attributes} & 45 (16.9\%) & 135 (50.6\%) & 45 (16.9\%) & 51 (19.1\%) & N/A & N/A \\ \midrule
\textbf{Functions} & 66 (24.7\%) & 134 (50.2\%) & 60 (22.5\%) & 58 (21.7\%) & 68 (25.5\%) & 67 (25.1\%) \\ \midrule
\textbf{Classes} & 35 (13.1\%) & 180 (67.4\%) & 26 (9.7\%) & 31 (11.6\%) & N/A & N/A \\ \bottomrule
\end{tabular}
}
\end{table}
%\decker{What is the overall accuracy not by type?  Is Stanford still the best, the full patterns seem to say no} \emily{So interesting the differences between individual tags \& identifiers as a whole! Pleasantly surprised how well Swum holds up in~\ref{tab:accuracytable}.)}

With this context in mind, we will now look at the agreement between the tagger-annotated and human-annotated grammar patterns (i.e., identifier-level accuracy analysis). This is shown in Table~\ref{tab:accuracytable}. We broke Stanford down into several columns in this table to see how its accuracy changes when we configure it differently. The configurations are as follows: We add an \textit{I} before function names before applying Stanford tagger. Additionally, some types of verbs can be considered adjectives in different contexts. Thus, we test the accuracy of Stanford under either assumption; the verb being used as a verb or being used as an adjective. Details of both all configurations for Stanford are given in Section~\ref{grammarpatterndef}.

This data shows that Swum had the highest agreement with the human-provided annotations at the level of grammar patterns.  Stanford had the second-highest agreement on average when we assume its best configuration-- though, Posse has a higher agreement with the human annotations in the Classes category regardless of Stanford's configuration. Swum's accuracy ranged between 50.2\% and 67.4\% while Posse and Stanford's ranged between 13.1\% - 24.7\% and 9.7\% - 26.6\% respectively. The difference in the results between Tables \ref{tab:pertagagreement} and \ref{tab:accuracytable} are interesting in that Stanford has the best performance in Table \ref{tab:pertagagreement} but under-performs Swum by a large margin in Table \ref{tab:accuracytable}. The reason for this difference is the ubiquity of noun modifiers in identifier names. Even though Stanford is more accurate on a larger set of part-of-speech tag categories, it under-performs on noun modifiers compared to Swum (15.71\% accuracy for Stanford vs 94.01\% for Swum), which consistently annotates noun modifiers correctly. Noun modifier is the most frequent annotation (with a frequency of 1,604 per Table~\ref{tab:pertagagreement}); Swum gets 1508 of these correct while Stanford gets 252, meaning Stanford missed 1256 words that Swum got correct. If we combine this with the fact that Swum's performance on the second-most-common annotation, nouns, is much closer to Stanford's (85.54\% accuracy for Swum vs 93.25\% for Stanford) than Stanford's performance is to Swum's on noun modifiers, it makes sense that, when looking at accuracy on annotating full identifier name grammar patterns (i.e., Table \ref{tab:accuracytable}), Swum outperforms Stanford despite Stanford's high annotation accuracy on most other part-of-speech tag types. In short, Stanford is more likely to get the very common \textit{NM+ N} pattern incorrect due to mis-annotating NM compared to Swum, which will occasionally mis-annotate N, but not as frequently as Stanford will mis-annotate NM.

Using this data, we can also confirm that Stanford's accuracy is increased in method names when appending and \textit{I} to the beginning of the name. This causes it to more accurately identify verbs. Additionally, Stanford's accuracy increases in methods when the verb specializations discussed in Section~\ref{grammarpatterndef} are assumed to be verbs and increases in every other category when they are assumed to be noun modifiers.

To understand more about the differences in agreement between the humans and taggers, we identified which patterns were most frequently incorrectly generated for each tagger, in part to understand each tagger's weaknesses. These patterns represent paths toward significantly improving the accuracy of part-of-speech taggers for source code identifiers. Table~\ref{tab:incorrectpatterns} gives the top five most frequently mis-annotated grammar patterns per category for each tagger used in our study. 

To contextualize the details we discuss below, we quickly summarize some of the core problems found in the part-of-speech tagger output using the data in Table~\ref{tab:incorrectpatterns}. Posse has trouble generating noun phrase and verb phrase patterns (i.e., \textit{NM+ N} and \textit{V NM+ N} respectively), including plural noun phrases such as \textit{NM+ NPL}. The reason for this is that Posse does not generally identify noun modifiers in sequences greater than 1 (i.e., it may annotate \textit{NM N}, but never \textit{NM NM N}). Even on sequences with only one noun modifier, it tends to prefer annotating noun modifiers as a noun. This problem is more pronounced in Stanford, which generally missed more noun phrase and verb phrase patterns than Posse. This makes sense as Stanford did not annotate noun modifiers very well (Table~\ref{tab:pertagagreement}). However, Stanford is very good at identifying plurals; Swum and Posse never identified plural words correctly in the dataset (Table~\ref{tab:pertagagreement}). This is why plural noun and verb phrase patterns are both frequently mis-annotated by both Swum and Posse. We will now focus our discussion around specific, commonly mis-annotated grammar patterns and discuss tagger annotations in the context of each.
%\emily{Seems like a hybrid POS approach where Stanford pre-annotates plurals, prepositions, etc \& passes the rest off to SWUM/POSSE type logic would do great!}

\begin{table}[]
\centering
\caption{Top 5 Most frequently mis-annotated grammar patterns}
\label{tab:incorrectpatterns}
\resizebox{!}{.28\paperheight}{%
\begin{tabular}{@{}llllll@{}}
\toprule
\multicolumn{6}{c}{\textbf{Attribute Names}} \\ \midrule
\multicolumn{2}{c|}{Posse} & \multicolumn{2}{c|}{Swum} & \multicolumn{2}{c|}{Stanford} \\ \midrule
NM N & \multicolumn{1}{l|}{58 (26.1\%)} & NM NPL & \multicolumn{1}{l|}{26 (19.7\%)} & NM N & 61 (28.4\%) \\ \midrule
NM NM N & \multicolumn{1}{l|}{31 (14.\%)} & NM NM NPL & \multicolumn{1}{l|}{9 (6.8\%)} & NM NM N & 33 (15.3\%) \\ \midrule
NM NPL & \multicolumn{1}{l|}{26 (11.7\%)} & NPL & \multicolumn{1}{l|}{9 (6.8\%)} & NM NPL & 19 (8.8\%) \\ \midrule
NM NM NM N & \multicolumn{1}{l|}{11 (5\%)} & PRE NM N & \multicolumn{1}{l|}{8 (6.1\%)} & NM NM NM N & 11 (5.1\%) \\ \midrule
NM NM NPL & \multicolumn{1}{l|}{9 (4.1\%)} & PRE N & \multicolumn{1}{l|}{7 (5.3\%)} & NM NM NPL & 9 (4.2\%) \\ \midrule
\multicolumn{6}{c}{\textbf{Declaration Names}} \\ \midrule
NM N & \multicolumn{1}{l|}{84 (38.2\%)} & NM NPL & \multicolumn{1}{l|}{30 (28.8\%)} & NM N & 88 (41.3\%) \\ \midrule
NM NM N & \multicolumn{1}{l|}{39 (17.7\%)} & NM NM NPL & \multicolumn{1}{l|}{8 (7.7\%)} & NM NM N & 42 (19.7\%) \\ \midrule
NM NPL & \multicolumn{1}{l|}{30 (13.6\%)} & N D & \multicolumn{1}{l|}{5 (4.8\%)} & NM NPL & 26 (12.2\%) \\ \midrule
NM NM NPL & \multicolumn{1}{l|}{8 (3.6\%)} & V N & \multicolumn{1}{l|}{5 (4.8\%)} & NM NM NPL & 8 (3.8\%) \\ \midrule
NM NM NM N & \multicolumn{1}{l|}{6 (2.7\%)} & N P N & \multicolumn{1}{l|}{4 (3.8\%)} & NM NM NM N & 6 (2.8\%) \\ \midrule
\multicolumn{6}{c}{\textbf{Parameter Names}} \\ \midrule
NM N & \multicolumn{1}{l|}{92 (44\%)} & NM NPL & \multicolumn{1}{l|}{20 (23.3\%)} & NM N & 92 (46.9\%) \\ \midrule
NM NM N & \multicolumn{1}{l|}{35 (16.7\%)} & NM NM NPL & \multicolumn{1}{l|}{12 (14.\%)} & NM NM N & 35 (17.9\%) \\ \midrule
NM NPL & \multicolumn{1}{l|}{20 (9.6\%)} & V N & \multicolumn{1}{l|}{6 (7\%)} & NM NPL & 15 (7.7\%) \\ \midrule
NM NM NPL & \multicolumn{1}{l|}{12 (5.7\%)} & NPL & \multicolumn{1}{l|}{5 (5.8\%)} & NM NM NPL & 12 (6.1\%) \\ \midrule
NPL & \multicolumn{1}{l|}{5 (2.4\%)} & NM N & \multicolumn{1}{l|}{3 (3.5\%)} & N & 6 (3.1\%) \\ \midrule
\multicolumn{6}{c}{\textbf{Function Names}} \\ \midrule
V NM N & \multicolumn{1}{l|}{33 (16.4\%)} & V NM NPL & \multicolumn{1}{l|}{10 (7.5\%)} & V NM N & 42 (21.1\%) \\ \midrule
V NM NM N & \multicolumn{1}{l|}{15 (7.5\%)} & NM N & \multicolumn{1}{l|}{6 (4.5\%)} & V NM NM N & 17 (8.5\%) \\ \midrule
V NM NPL & \multicolumn{1}{l|}{10 (5\%)} & V NM NM NPL & \multicolumn{1}{l|}{6 (4.5\%)} & NM NM N & 10 (5\%) \\ \midrule
NM NM N & \multicolumn{1}{l|}{10 (5\%)} & N V & \multicolumn{1}{l|}{5 (3.8\%)} & NM N & 8 (4\%) \\ \midrule
NM N & \multicolumn{1}{l|}{9 (4.5\%)} & V NPL & \multicolumn{1}{l|}{4 (3\%)} & V NM NPL & 8 (4\%) \\ \midrule
\multicolumn{6}{c}{\textbf{Class Names}} \\ \midrule
NM NM N & \multicolumn{1}{l|}{70 (30.2\%)} & PRE NM N & \multicolumn{1}{l|}{10 (11.5\%)} & NM N & 72 (30.5\%) \\ \midrule
NM N & \multicolumn{1}{l|}{63 (27.2\%)} & NM NPL & \multicolumn{1}{l|}{8 (9.2\%)} & NM NM N & 69 (29.2\%) \\ \midrule
NM NM NM N & \multicolumn{1}{l|}{16 (6.9\%)} & NM N NM & \multicolumn{1}{l|}{7 (8\%)} & NM NM NM N & 16 (6.8\%) \\ \midrule
PRE NM N & \multicolumn{1}{l|}{10 (4.3\%)} & NM NM N & \multicolumn{1}{l|}{5 (5.7\%)} & PRE NM N & 10 (4.2\%) \\ \midrule
NM NPL & \multicolumn{1}{l|}{8 (3.4\%)} & PRE NM NM N & \multicolumn{1}{l|}{5 (5.7\%)} & NM N NM & 7 (3\%) \\ \bottomrule
\end{tabular}
}
\end{table}

\textbf{Grammar Pattern \textit{NM+ N}:} Swum was the best tagger at identifying noun phrase patterns because it very accurately recognized noun modifiers. It did overestimate noun phrase patterns due to over-annotating noun modifiers where they do not belong. \textbf{Example}: \textit{rotation Per Second} has a pattern \textit{N P N}. Swum mis-annotates two out of three words by annotating this identifier as \textit{NM NM N}; failing to recognize the fact that \textit{Per} is a preposition in this context. 

Posse and Stanford have a harder time with noun phrase patterns and tend to use a noun instead of a noun modifier. \textbf{Example}: \textit{cache entity} and \textit{root ptr} are both given an \textit{N N} pattern by Stanford and Posse. While annotating using noun is not wholly inappropriate, these nouns play a double role of both identifying an external concept which exists as its own object (the root, the cache; both of which are nouns) and using this concept to modify the head-noun-of-interest (e.g., the entity, the pointer) so the developer fully understands what they are dealing with; root and cache are nouns being used as noun modifiers and not pure nouns. An easy way to see this adjectival relationship is to add a dash between the words; root-pointer, cache-entity. Root describes the pointer, cache describes the entity.

\textbf{Grammar Pattern \textit{NM* NPL}:}  Posse and Swum do not detect plurals, while Stanford is very good at detecting them (see Table~\ref{tab:pertagagreement}). Stanford tends to get noun plurals individually correct even when it would mistakenly annotate a noun modifier as a noun, which is why \textit{NM NPL} was still one of the most common patterns for Stanford to mis-annotate. \textbf{Example}: \textit{num active contexts} has a pattern \textit{NM NM NPL} due to the plural at the end. Swum does not recognize the plural and gives it a \textit{NM NM N} pattern. Posse gave it an \textit{N NM N} pattern and Stanford gave it a \textit{N NM NPL} pattern. Thus, Stanford and Swum were both nearest to the correct solution despite both being incorrect. Stanford did not have trouble with \textit{NPL} patterns when there were no noun modifiers. This suggests a very good way that tagger annotations may be combined; Stanford can identify noun plurals accurately and Swum can identify noun-modifiers accurately. In general, this was the hardest, frequently observed pattern for any individual tagger to annotate completely correctly.

\textbf{Grammar Pattern \textit{V NM+ N}:} The deciding factor in mis-annotating verb phrase patterns tended to be noun modifiers and plural nouns. Generally speaking, the taggers agreed with the human annotators on the position of the verb in method names between 56\% and 76\% of the time. This contrasts with how often they agreed on the full human annotation in function names (Table~\ref{tab:accuracytable}); 24.7\% of the time for Posse, 50.2\% of the time for Swum, 25.5\% of the time for Stanford. All taggers still have problems determining the correct verb, but detecting noun modifiers is the bigger issue. 

Posse and Stanford had the most trouble with this pattern; it is not in Swum's top 5. \textbf{Examples}: \textit{reset meta class cache} and \textit{set project naming strategy}; both with a human-annotated grammar pattern of \textit{V NM NM N}. Swum agreed with the human annotators on both; Stanford annotated both with \textit{V N N N}; and Posse annotated these as \textit{V N N N} and \textit{V N NM N} respectively.

%While Swum did well in most cases, it did not always annotate this pattern correctly. For example, it \textit{check GL Support} as \textit{V V N}; potentially due to having trouble with the abbreviation.

\textbf{Grammar Pattern \textit{V N}:} This pattern tends to be mis-annotated when the part-of-speech taggers could not determine which (if any) word in the identifier was a verb. One of the most common situations for this was identifiers with a boolean type. Posse and Swum tend to expect that there is a verb when they know that they are looking at a function name, but in non-function identifiers they are less likely annotate using verb. For example, the \textit{write root} identifier has a human-annotated pattern of \textit{V N}. Stanford agrees with the human-annotated pattern, but Swum mis-annotates it as \textit{NM N} and Posse as \textit{N N}.  

\textbf{Grammar Pattern \textit{V NM* NPL}:} This pattern is similar to \textit{V N} in that one of the biggest problems the taggers had was annotating the verb. However, this pattern was more trouble for Posse and Swum than Stanford due to the inclusion of a plural-- Stanford detects plurals well, but neither Swum or Posse are able to determine when a word is in a plural form.

\textbf{Grammar Pattern \textit{PRE...}:} We will discuss all patterns with a preamble here (i.e., the ... could be any other pattern, such as noun phrase or verb phrase). Preambles were difficult for all taggers to deal with. Swum rarely detects preambles while Posse and Stanford do not detect them at all. Generally, a preamble is mis-annotated as noun for Posse and Stanford or noun modifier for Swum. The problem with detecting preambles is that some prior information is required-- a tagger needs to scan the code and/or be able to recognize naming conventions such as Hungarian \cite{hungariannotation}, to determine which character sequences are being used as preambles.  There are also cases where it is not clear that a frequent character sequence is a preamble. \textbf{Examples}: the GRPC project tends to append \textit{grpc} before many of its identifiers-- \textit{grpc json writer value string} has a grammar pattern of \textit{PRE NM NM NM N}. A scan of GRPCs code could identify this as a preamble. \textit{XML} is sometimes used frequently at the beginning of function names such as \textit{xmlWriter, xmlReader}. It may look like a preamble in some systems due to how frequently it appears at the beginning of an identifier, but is not a preamble (Section \ref{grammarpatterndef}) because it specializes our understanding of the words \textit{reader} and \textit{writer}. This suggests that some domain knowledge is additionally required to correctly determine when a character sequence is a preamble.

\textbf{Grammar Pattern \textit{N P N}:} This is a prepositional phrase grammar pattern. Swum and Posse had difficulty identifying prepositions while Stanford was effective at it. Stanford tends to do well on this pattern; it usually annotated the preposition correctly but would occasionally mistake noun for verb. This pattern is one of the less common ones in our dataset, but is an example of a pattern on which Stanford is more accurate than Swum or Posse.

\textbf{Grammar Pattern \textit{N D}:} This pattern is a noun followed by a number. Swum and Posse cannot detect digits, thus Stanford is the only tagger that correctly identifies this pattern. Stanford had high accuracy on digits; agreeing with all digits identified by human annotators. It occasionally is unable to annotate the noun correctly; many of these cases are when there is an abbreviation or a word which is colloquial (or domain-specific) to programmers.

%\textbf{Grammar Pattern \textit{V} and Grammar Pattern \textit{N}:} These are the two smallest patterns and correctly annotating either one depending primarily on two things: 1) the tagger recognized the word. 2) the tagger is effective at guessing the tag for words which it does not recognize. 

\textbf{\textit{{Summary for RQ2}}}: The highest amount of agreement was between Swum and the human annotators; Swum's accuracy ranged between 50.2\% and 67.8\% while Posse and Stanford's ranged between 13.1\% - 24.7\% and 9.7\% - 26.6\% respectively. The most frequently incorrectly annotated patterns were: 1) singular noun phrases for Stanford, plural noun phrases for Swum and both for Posse. 2) plural verb phrases for Swum, singular verb phrases for Stanford, and both for Posse. 3) grammar patterns which include a preamble for all three taggers. Our results 1) indicate that it is possible to correctly annotate abbreviations without expanding them in some cases, particularly in noun phrases; this is supported by Swum's accuracy. 2) indicate that these taggers have complementary strengths and weaknesses, meaning that their output can be combined into a more accurate result than they are able to obtain individually. As a simple example, Stanford annotates noun plurals fairly accurately; this could be used to improve Swum or Posse's output on \textit{NM NPL}, while Swum/Posse can improve Stanford's \textit{NM} detection; causing all three taggers to get \textit{NM NPL} correct more frequently. 3) confirm that Stanford's accuracy on functions is improved by adding \textit{I} and that certain verb forms are more likely to be verbs when found in function names but adjectives when found in other types of identifiers. This also shows that Swum and Posse need to detect specialized verb forms in order to correctly identify when these verbs are used as verbs or adjectives. 4) show that code context and domain-specific information are very important for annotating words correctly; preambles are one of the categories that would most benefit from taggers which are able to leverage surrounding code structure, naming conventions, and domain information.

\subsection{RQ3: \RQC}\label{uncommonpatterns}
We observed a number of grammar patterns that were not frequent enough to be in the top 5, but nevertheless appear in multiple systems. The question is: What are these patterns? What types of identifiers do they represent? To answer these questions, we manually looked through the set of human-annotated grammar patterns and picked patterns which occurred two or more times in any single category (i.e., function, class, etc.) and are not similar to the patterns we discussed in RQ1. This resulted in the following grammar patterns:

\textbf{Grammar Pattern \textit{NM* N P N}:} This pattern is a noun phrase and a prepositional phrase combined. \textbf{Examples}: \textit{depth stencil as texture} and \textit{scroll id for node}. Identifiers with this pattern describe the relationship between a noun phrase on the left of the preposition and a noun phrase on the right. The noun phrase on the right and left both contain a head-noun. In this case, \textit{stencil} and \textit{id} are the left head-nouns (lhn) while \textit{texture} and \textit{node} are the right head-nouns (rhn). The preposition tells us how these head-nouns are related to one another and how this relationship defines the identifier. In \textit{depth stencil as texture} we are told that this identifier represents the texture version of a stencil-- or a conversion from stencil to texture. In \textit{scroll id for node} we are told that this identifier represents an id for a specific node. 

An example which does not include a noun modifier is \textit{angle in radian} with a grammar pattern of \textit{N P N}. The same concepts above apply-- \textit{angle} is the lhn and \textit{radian} is the rhn. The preposition \textit{in} helps us understand how their relationship defines this identifier. In this case, the identifier represents an angle using radians as the unit of measurement. 

\textbf{Grammar Pattern \textit{P N}:} This pattern is a preposition followed by a noun. An example of this pattern is the identifier \textit{on connect}, which specifies an event to be fired when a connection is made. Other instances of this pattern include \textit{with charset} and \textit{from server}. The former is a function which takes a charset as its parameter, and the latter is a boolean which tells the developer whether certain data was obtained from a server. Identifiers in event-driven systems likely use this pattern, or its derivatives, often (e.g., onButtonPress, onEnter). This suggests that more unique grammar patterns may be obtained by studying identifiers found in systems using certain architectural, or programming, patterns.

\textbf{Grammar Pattern \textit{DT NM* NPL:}} this pattern is a determiner followed by a plural noun phrase. \textbf{Example}: \textit{all invocation matchers}. Determiners like \textit{all} are called quantifiers. They indicate how much or how little of the head-noun is being represented. So in this case, the identifier represents a list of every invocation matcher. This pattern is familiar in that it contains a plural noun phrase, but the inclusion of the determiner quantifies the plural noun phrase more formally than if it did not include it; \textit{invocation matchers} without \textit{all} indicates a list of invocation matchers, but \textit{all invocation matchers} tells us the specific population matchers included in the list. The word \textit{all} was the most common determiner used for the identifiers that fit this pattern in our dataset. This pattern may show up more in code that deals with querying-- databases or other query-able structures, where words like \textit{all} and \textit{any} might be useful. Further study is required to determine common contexts for determiners in source code.

\textbf{Grammar Pattern \textit{V+}:} Like the other patterns derived from verbs above, this one is typically used with Boolean variables and functions. One interesting thing about this pattern is that there there is no noun for the verb to act on. When this happens in a function name, it is because the noun is contained within the function's arguments or is the calling object itself (i.e., this). \textbf{Examples}: \textit{delete}, \textit{do forward}, \textit{parsing}, and \textit{sort}. A delete function could either be applied to an argument to \textit{delete} that argument, or to the calling object to delete some internal memory. The \textit{do forward} function in our dataset redirects a user (i.e., \textit{forward} is being used as synonym for the verb \textit{redirect}), and the system it is from uses \textit{do} to prefix methods which perform, for example, HTTP actions. The V+ pattern can also represent Boolean variables that are not function names. The \textit{parsing} identifier is a Boolean variable in our dataset which is annotated as verb since it is asking a true or false in reference to an action-- specifically, a parsing action. \textit{Sort} represents a function where the user can supply a predicate to influence how elements are sorted by the function. This pattern may appear more often in generic libraries, where the head-noun is not supplied by the library creators due to the generic nature of the solution. Instead, the user will supply a head-noun when they use the library.

\textbf{Grammar Pattern \textit{V P NM N:}} This pattern is a verb followed by a prepositional phrase. This was found only among function identifier names. \textbf{Examples}: \textit{convert to php namespace} and \textit{register with volatility spread}. This pattern uses a verb to specify an action on an unknown entity (e.g., the identifiers above do not specify what to \textit{convert} or what to \textit{register}) and uses the noun phrase on the right side of the preposition as a reference point; the specific thing to which this unknown entity will be related. The nature of this relationship is specified by the verb and preposition (i.e., convert to, register with).

While there are other grammar patterns which we do not discuss, many of them occur only once or are very similar in structure to grammar patterns that we have already discussed above.

\textbf{\textit{{Summary for RQ3}}}: There are many ways to describe interesting types of program behavior and semantics. In RQ3 we have discussed less frequent, yet legitimate, grammar patterns. This adds diversity to the group of grammar patterns discussed previously, and highlights the need for further research to uncover new grammar patterns. This will help ensure that we obtain an understanding of both the breadth and depth of grammar patterns used to describe different forms of program semantics and behavior. Grammar patterns which include prepositional phrases and determiners are less frequent than other patterns in our dataset. Yet, as we have pointed out, some of these patterns are copacetic with certain domains. Prepositional phrase grammar patterns are used in event-driven programming very frequently, for example. The patterns presented in RQ3 represent future directions for research; there are not enough examples of the patterns we presented in this research question to draw strong conclusions, but their existence is evidence that these grammar patterns may be common in other contexts (e.g., different architecture and design patterns). We argue that studying grammar patterns in these other contexts will result in more semantics belying those patterns than what we have discussed, or perhaps even new patterns. These domain-specific patterns would be very useful for suggesting and appraising identifier names within those contexts.

\begin{table}[]
\centering
\caption{Grammar patterns broken down by language}
\label{tab:languagebasedgrammarpatterns}
\resizebox{!}{.1\paperheight}{%
\begin{tabular}{@{}ll|ll|ll@{}}
\toprule
\multicolumn{2}{c|}{\textbf{C Language Patterns}} & \multicolumn{2}{c|}{\textbf{C++ Patterns}} & \multicolumn{2}{c|}{\textbf{Java Patterns}} \\ \midrule
NM N & 76 (33\%) & NM N & 175 (30.5\%) & NM N & 154 (30.1\%) \\ \midrule
NM NM N & 17 (7.4\%) & NM NM N & 93 (16.2\%) & NM NM N & 81 (15.8\%) \\ \midrule
NM NPL & 14 (6.1\%) & NM NPL & 31 (5.4\%) & NM NPL & 42 (8.2\%) \\ \midrule
N & 11 (4.8\%) & N & 28 (4.9\%) & V NM N & 21 (4.1\%) \\ \midrule
V N & 9 (3.9\%) & V N & 27 (4.7\%) & NM NM NPL & 20 (3.9\%) \\ \midrule
V NM N & 7 (3\%) & V NM N & 25 (4.4\%) & NM NM NM N & 18 (3.5\%) \\ \midrule
NM NM NPL & 6 (2.6\%) & NM NM NM N & 14 (2.4\%) & N & 16 (3.1\%) \\ \midrule
NM NM NM N & 5 (2.2\%) & PRE NM N & 12 (2.1\%) & V NM NPL & 10 (2\%) \\ \midrule
NM V NM N & 4 (1.7\%) & V NM NM N & 9 (1.6\%) & V NM NM N & 9 (1.8\%) \\ \midrule
PRE NM N V N & 3 (1.3\%) & NPL & 9 (1.6\%) & PRE NM N & 7 (1.4\%) \\ \bottomrule
\end{tabular}
}
\end{table}

\subsection{RQ4: \RQD}\label{languagepatterns}
Programming languages have their own individual characteristics. To give a few (i.e., non-exhaustive) examples: C is a procedural language that does not support objected oriented programming, Java is an object oriented language, and C++ supports facets of object orientated programming and generic programming. Given this, we grouped grammar patterns in our data set by programming language with a goal varying the language to see if certain grammar patterns were more common to a specific language. This data is shown in Table \ref{tab:languagebasedgrammarpatterns}, where we show the top 10 patterns C, C++, and Java. We note that most of the identifiers in our set were either C++ (573) or Java (511) identifiers; a smaller number were C (229) and C-sharp (22). We leave C-sharp out of our analysis due to the very small number of them, but still make this data available in our open dataset \footnote{https://scanl.org/}. We include C in our analysis, but note that the results for C may not generalize as well as for C++ and Java. 

The results in this table show that the identifiers found in individual languages are largely similar to one another. Most patterns that occurred more than once in any language also occurred in the other languages. To get a better look at language-specific patterns, we looked for any pattern which occurred multiple times but only in a specific language. After finding these patterns, we manually examined and picked patterns which were not reflective of system-specific naming conventions (i.e., only occurs in a single system). In general, the language-specific patterns tended to include determiners (DT), prepositions (P), or digits (D). For example, identifiers with grammar patterns including these annotations are: \textit{all action roots (DT NM* N, Java)}, \textit{group by context (N P N, Java)}, and \textit{event 0 (N D, C++)}. We discussed the former two patterns in RQ3. The last pattern, \textit{N D} is used typically as a way to distinguish two identifiers which otherwise have the same name (i.e., event0, event1, etc). We did not find any patterns which were significantly programming language-specific. 

However, this result is not a definitive answer on the differences (or lack thereof) in grammar patterns between programming languages. While it does show us that there is a lot of similarity in grammar patterns between languages, another way to interpret this data is that these differences are unlikely to be found without controlling for other factors. For example, the programming paradigms, architectural/design patterns, and problem domain of the systems in the dataset. If controlled for, these factors could reveal differences in the grammar patterns between different programming languages.

\begin{table}[]
\centering
\caption{Distribution of abbreviations and dictionary terms between different languages in the data set}
\label{tab:abbreviation_distribution}
\resizebox{!}{.052\paperheight}{%
\begin{tabular}{@{}lllllc@{}}
\toprule
\multicolumn{1}{c}{\textbf{Word Type}} & \multicolumn{1}{c}{\textbf{C}} & \multicolumn{1}{c}{\textbf{C++}} & \multicolumn{1}{c}{\textbf{Java}} & \textbf{C\#} & \multicolumn{1}{l}{\textbf{\begin{tabular}[c]{@{}l@{}}Total in \\ Dataset\end{tabular}}} \\ \midrule
Abbreviations & 114 (22.6\%) & 223 (17.4\%) & 173 (14.5\%) & 9 (17.3\%) & 519 \\ \midrule
Dictionary Terms & 505 & 1282 & 1192 & 52 & 3031 \\ \bottomrule
\end{tabular}
}
\end{table}

\begin{table}[]
\centering
\caption{Accuracy of taggers on abbreviated and non-abbreviated terms}
\label{tab:abbreviation_accuracy}
\begin{tabular}{@{}llllc@{}}
\toprule
\multicolumn{1}{c}{\textbf{Word Type}} & \multicolumn{1}{c}{\textbf{Posse}} & \multicolumn{1}{c}{\textbf{Swum}} & \multicolumn{1}{c}{\textbf{Stanford}} & \multicolumn{1}{l}{\textbf{\begin{tabular}[c]{@{}l@{}}Total in \\ Dataset\end{tabular}}} \\ \midrule
Abbreviations & 183 (35.3\%) & 345 (66.5\%) & 230 (44.3\%) & 519 \\ \midrule
Dictionary Terms & 1484 (49\%) & 2321 (76.6\%) & 1624 (53.6\%) & 3031 \\ \bottomrule
\end{tabular}
\end{table}
We then looked at the distribution of abbreviations and dictionary words between Java and C/C++ to provide more insight about the differences in identifier structure between languages. Since abbreviations may make it difficult to obtain the meaning of a word, part-of-speech taggers might annotate these words less accurately. Table~\ref{tab:abbreviation_distribution} shows the distribution. We include C\# in this table for completeness despite its low number of identifiers. To determine if a token is a full word or an abbreviation, we used Wordnet \cite{miller1995wordnet}. If Wordnet recognized the word, we considered it a full word. Otherwise, it is an abbreviation. The results indicate that C and C++ tend to contain more abbreviated terms. 

Given this, we then looked at the accuracy of each part-of-speech tagger on identifiers from different languages. This data is found in Table~\ref{tab:abbreviation_accuracy}. All taggers had decreased performance on abbreviated terms, but Posse was the least accurate and saw the most significant decrease in performance (-14\% from its full word accuracy). Finally, given this data about tagger performance on abbreviations and the distribution of abbreviations in different languages, we looked at tagger accuracy per programming language. This data is in Table~\ref{tab:languageaccuracytable}. While Posse/Stanford performed better on Java than C/C++ systems, their performance degraded a total of 3.4\% and 1.4\% respectively. Swum, however, was nearly 12\% less accurate on C identifiers compared to C++ identifiers. Swum also performed better on C++ identifiers versus Java, unlike the other two taggers.

\begin{table}[]
\centering
\caption{Accuracy of part-of-speech taggers split by programming language}
\label{tab:languageaccuracytable}
\begin{tabular}{@{}cccc@{}}
\toprule
\textbf{} & \textbf{Posse} & \textbf{Swum} & \textbf{Stanford} \\ \midrule
\textbf{C} & 37 (16.2\%) & 116 (50.7\%) & 45 (19.7\%) \\ \midrule
\textbf{C++} & 107 (18.7\%) & 357 (62.3\%) & 116 (20.2\%) \\ \midrule
\textbf{Java} & 100 (19.6\%) & 305 (59.7\%) & 108 (21.1\%) \\ \bottomrule
\end{tabular}
\end{table}

\textbf{\textit{{Summary for RQ4}}}: At the level of grammar patterns, while only controlling for identifier category (e.g., function name) and programming language, there does not appear to be a significant difference in the grammar patterns for Java and C/C++ identifiers. It may be the case that significant grammar pattern differences appear when controlling for more confounding factors and we believe this would be a strong direction for future work. We do note a difference in the use of abbreviations between C/C++ and Java identifiers, where Java identifiers tend to have fewer abbreviations than the former two languages. Abbreviations can hinder the accuracy of part-of-speech taggers, which we confirmed by examining the annotations given to abbreviations by the three taggers in this study; all taggers performed worse on abbreviations than on full words. However, difficulty with abbreviations did not significantly reduce Posse/Stanford's performance between programming languages, indicating that abbreviations are not the biggest problem these taggers face, while expanding abbreviations may significantly (up to 10\%) improve Swum's performance.

\section{THREATS TO VALIDITY} \label{threats}
This study was done on a set of 1,335 identifiers; the largest set of open-source, manually tagged, cross-language identifiers at the time of writing. Even so, there is a threat that the annotated set contains imperfections. To mitigate this, we used cross-validation, where all annotators performed a validation step on all grammar patterns; each grammar pattern was validated by two annotators beside the original annotator. Additionally, the dataset is publicly available; future corrections are possible. We calculated that a statistically representative sample for the size of our dataset of 20 systems is 267 given a 95\% confidence level and a confidence interval of 6\%. We picked 95\% and 6\% as a trade-off between representativeness of the sample, the amount of manual labor required of the annotators, and the sample size used in prior studies.

In this study, we intended to focus only on production code. Thus, we have excluded test files. Yet, if developers violate JUnit test conventions by including non-annotated test functions in their production code, we may have picked these identifiers to be included in the dataset. This threat is mitigated by the fact that we manually examined every identifier.

We did not expand abbreviations and did not remove identifiers that contained abbreviations or non-dictionary terms. This resulted in an increased number of mistakes made by the taggers, as shown in RQ4. Abbreviations were included so we could measure the accuracy of part-of-speech taggers in a real-world environment, where abbreviation expansion might not be feasible. To help ensure that the humans did not mistake the annotation for abbreviated terms, the human annotators were allowed to look at the code. Additionally, we rely on identifier splitting techniques. While automated splitting is highly accurate, it is not perfect; we mitigated this problem by manually examining every identifier and correcting the split where there were mistakes.

We sampled identifiers from multiple programming languages to avoid biasing the names in our identifier set toward a particular programming language. While this means that the set we obtain is more likely to be generalizable, it also means that particular language-specific patterns may not be detected. To mitigate this, we present frequent patterns per language (Java, C, and C++). However, we had fewer identifiers in C than in Java and C++, meaning that the results for C may not be as generalizable as for C++ and Java. In addition, it is not always possible to identify an identifier as being a C or C++ identifier. In this study, we assume .c and .h files are for C systems while .cpp and .hpp files are for C++ systems. However, this only a naming convention; C++ programs may use .c and .h, for example. Next, while C, C++ and Java are different programming languages, they are all in similar programming paradigms (e.g., imperative languages). Our results may not extend to languages in other paradigms, such as functional languages. Finally, our analysis of the differences between grammar patterns in different programming languages does not take into account several potential confounding factors including system domain and design patterns. Thus, we acknowledge this problem and recommend directions for future research, based on our data, to mitigate this threat.

\section{RELATED WORK}
\label{sec:related}

\subsection{Rename Analysis}
Arnoudova et al. \cite{Arnaoudova:2014} present an approach to analyze and classify identifier renamings. The authors show the impact of proper naming on minimizing software development effort and find that 68\% of developers think recommending identifier names would be useful. They also defined a catalog of linguistic anti-patterns \cite{Arnaoudova:2013}.  Liu et al.\cite{liu2015identifying} proposed an approach that recommends a batch of rename operations to code elements closely related to the rename. They also studied the relationship between argument and parameter names to detect naming anomalies and suggest renames \cite{liu2016nomen}. Peruma et al. \cite{Peruma:2018:EIW:3242163.3242169} studied how terms in an identifier change and contextualized these changes by analyzing commit messages using a topic modeler. They later extend this work to include refactorings \cite{Perumascam} and data type changes \cite{PERUMA2020110704} that co-occur with renames.

These techniques are concerned with examining the structure and semantics of names as they evolve through renames. By contrast, we present the structure and semantics of names as they stand at a single point in the version history of a set of systems. Rename analysis and our work are complimentary; our analysis of naming structure can be used to help improve how these techniques analyze changes between two versions of a name, in part by helping to improve off-the-shelf part-of-speech taggers. For example, Arnaoudova's \cite{Arnaoudova:2014} work depends on Wordnet and Stanford; we have shown that Stanford needs significant help to accurately annotate identifiers. Beyond part-of-speech tagging, our work also highlights many different grammar patterns. These patterns can be used to improve our understanding of how names evolve over time by examining how the patterns evolve.

\subsection{Identifier Type and Name Generation}
There are several recent approaches to appraising identifier names for variables, functions, and classes. Kashiwabara et al. \cite{Kashiwabara:2014} use association rule mining to identify verbs that might be good candidates for use in method names. Abebe \cite{Abebe:2013} uses an ontology that models the word relationships within a piece of software. Allamanis et al. \cite{Allamanis:2015} introduce a novel language model called the Subtoken Context Model. One thing these approaches have in common is the use of frequent tokens and source code context to try and generate high-quality identifier names. In contrast, in this paper, we manually analyze identifier names to understand their structure and how that structure affects the role of individual words within the structure. A better understanding of identifier structure in different contexts can synergize with automated techniques like those above by helping them generate structurally consistent identifiers. One thing that automated techniques struggle with is generating identifiers with words which are not present in the surrounding code (i.e., neologisms \cite{Allamanis:2015}); a stronger understanding of naming structure combined with better part-of-speech tagging and word relationships in software (e.g., \cite{falleriWordnet}) will allow these techniques to generate higher-quality names by providing guidance on what grammatical structure a new identifier should use so that new words can be picked to satisfy that structure. A stronger understanding of grammatical structure in identifiers can also help researchers train their approaches more effectively.

There has also been work in reverse engineer data types from identifiers \cite{malik2019,Hellendoorn:2018}, focusing on using identifier names in languages such as Javascript to determine the type of the identifier. This relates most closely to our observations about List and Boolean identifiers. For example, Malik et al \cite{malik2019} observe that some identifiers are more likely to give away their type and their example is boolean functions, which commonly start with the words "is" or "has" and thus can be inferred to be boolean. The primary difference in our work is that the type is already available in our chosen languages (whereas they focus on Javascript), so we do not need to infer it. Instead, we are more interested in how certain aspects of the type name influences the structure of the corresponding identifier name and how we can use that relationship to improve the identifier. Type inference work like those above do indicate the presence of type clues within the identifier name or its code context. There is some potential for synergy in their work and ours; their approaches highlight how to find type clues in the source code while some of the data in our study highlights where name suggestions could be used to improve identifier naming practices (e.g., in identifiers with collection and boolean types). Improving these practices might make natural-language-based type inference approaches more accurate and efficient, since identifier name structure will be more consistent and well-defined. In addition, examining the data from these type inference approaches might point out some other ways to improve identifier naming appraisal or suggestions using more source code context to better determine when, for example, to use a verb in a boolean or a plural for a list identifier.

\subsection{Software Ontology Creation Using Identifier Names}

A lot of work has been done in the area of modeling domain knowledge and word relationships by leveraging identifiers \cite{AbebeDomainConcepts, RatiuProgramsKnowledgeBases, RatiuConceptMapping, Deissenboeck06aunified, falleriWordnet}. Abebe and Tonella \cite{AbebeDomainConcepts} analyze the effectiveness of information retrieval based techniques for filtering domain concepts and relations from implementation details. They show that fully automated techniques based on keywords or topics have low performance but that a semi-automated approach  can significantly improve results. Falleri et al., present a way to automatically construct a wordnet-like \cite{miller1995wordnet} identifier network from software. Their model is based on synonymy, hypernymy and hyponymy, which are types of relationships between words. Synonyms are words with similar or equivalent meaning; hyper/hyponyms are words which, relative to one another, have a broader or more narrow domain (e.g., dog is a hyponym of animal, animal is a hypernym of dog). Ratiu and Deissenboeck \cite{RatiuConceptMapping} present a framework for mapping real world concepts to program elements bi-directionally. They use a set of object-oriented properties (e.g., isA, hasA) to map relationships between program elements and string matching to map these elements to external concepts. This extends two prior works of theirs; one paper on a previous version of their meta model \cite{Deissenboeck06aunified} and a second paper on linking programs to ontologies \cite{RatiuProgramsKnowledgeBases}. Many of these approaches need to split and analyze words found in an identifier in order to connect these identifiers to a model of program semantics (e.g., class hierarchies). All of these approaches rely on identifiers.

While we do not construct an ontology in this paper, many software word ontologies use meta-data about words to understand the relationship between different words. There is a synergistic relationship between the work we present here and software ontologies since stronger ontologies can help us generate and study grammar patterns effectively (e.g., by increasing our knowledge of word meaning/relationships) and the data used in our paper can help construct stronger software word ontologies by supporting research in part of speech taggers for source code and providing annotated word datasets from which word relationships can be deduced. In the future, we aim to use the identifiers and grammar patterns in our dataset to improve software word ontologies.

\subsection{Identifier Structure and Semantics Analysis}

Liblit et al.~\cite{Liblit06cognitiveperspectives} discusses naming in several programming languages and makes observations about how natural language influences the use of words in these languages. Schankin et al. \cite{Schankin:2018} focus on investigating the impact of more informative identifiers on code comprehension. Their findings show an advantage of descriptive identifiers over non-descriptive ones. Hofmeister et al \cite{Hofmeister:2017} compared comprehension of identifiers containing words against identifiers containing letters and/or abbreviations. Their results show that when identifiers contained only words instead of abbreviations or letters, developer comprehension speed increased by 19\% on average. Lawrie et al \cite{Binkley2006} did a study and used three different "levels" of identifiers. The results show that full word identifiers lead to the best comprehension compared to the other levels studied. Butler's work ~\cite{butler2010exploring} extends their previous work on Java class identifiers \cite{Butler:2009} to show that flawed method identifiers are also associated with low-quality code according to static analysis-based metrics. These papers primarily study the words found in identifiers and how they relate to code behavior or comprehension rather than word metadata (e.g., part-of-speech).

Caprile and Tonella \cite{tonella:1999} analyzes the syntax and semantics of function identifiers. They create classes which can be used to understand the behavior of a function; grouping function identifiers by leveraging the words within them to understand some of the semantics of those identifiers. While they do not identify particular grammar patterns, this study does identify grammatical elements in function identifiers, such as noun and verb, and discusses different roles that they play in expressing behavior both independently and in conjunction using the classes they propose. They also use the classes identified in this prior work to propose methods for restructuring program identifiers \cite{tonella:2000}. Fry and Shepherd \cite{Shepherd:2007,fry:2008} study verb-direct objects to link verbs to the natural-language-representation of the entity they act upon in order to assist in locating action-oriented concerns. The primary concern in this work is identifying the entity (e.g., an object) which a verb is targeting (e.g., the action part of a method name). We study a broader set of identifiers and focus on grammar patterns in general, rather than those containing (or related to) a verb. The grammar structures we prevent may help in identifying verb-direct objects.

H{\o}st and {\O}stvold study method names as part of a line of work discussed in H{\o}st's disseration \cite{hostdissertation}. This line of work starts by analyzing a corpus of Java method implementations to establish the meanings of verbs in method names based on method behavior, which they measure using a set of attributes which they define \cite{HostLexicon}. They automatically create a lexicon of verbs that are commonly used by developers and a way to compare verbs in this lexicon by analyzing their program semantics. They build on this work in \cite{hostphrasebook} by using full method names which they refer to as phrases and augment their semantic model by considering a richer set of attributes. The outcome of this work is that they were able to aggregate methods by their phrases and come up with the semantics behind those phrases using their semantic model, therefore modeling the relationship between method names and method behavior. The phrases they discuss are very similar to the grammar patterns we present. They extend this use of phrases by presenting an approach to debug method names \cite{Host:2009}. In this work, they designed automated naming rules using method signature elements. They use the phrase refinement from their prior paper, which takes a sequence of part-of-speech tags (i.e., phrases) and concretes them by substituting real words. (e.g., the phrase $<$verb$>$-$<$adjective$>$ might refine to is-empty). They connect these patterns to different method behaviors and use this to determine when a method's name and implementation do not match. They consider this a naming bug. Finally, in \cite{Hst2011CanonicalMN}, H{\o}st and {\O}stvold analyzed how ambiguous verbs in method names makes comprehension of Java programs more difficult. They proposed a way to detect when two or more verbs are synonymous and being used to describe the same behavior in a program; hoping to eliminate these redundancies as well as increase naming consistency and correctness. They perform this detection using two metrics which they introduce called nominal and semantic entropy. H{\o}st and {\O}stvold's work focuses heavily on method naming patterns; connecting these to the implementation of the method to both understand and critique method naming. We focus on a larger variety of identifiers (i.e., not only method names) and have a different goal-- specifically exploring and analyzing the wide variety of grammar patterns in different types of identifiers and understanding the effectiveness of part-of-speech taggers. Our work can complement H{\o}st and {\O}stvold's.

Butler \cite{butler:2011} studied class identifier names and lexical inheritance; analyzing the effect that interfaces or inheritance has on the name of a given class. For example, a class may inherit from a super class or implement a particular interface. Sometimes this class will incorporate words from the interface name or inherited class in its name. His study builds on work by Singer and Kirkham \cite{singer:2008}, who identified a grammar pattern for class names of (adjective)* (noun)+ and studies how class names correlate with micro patterns. Amongst Butler’s findings, he identifies a number of grammar patterns for class names: (noun)+, (adjective)+ (noun)+, (noun)+ (adjective)+ (noun)+, (verb) (noun)+ and extends these patterns to identify where inherited names and interface names appear in the pattern. The same author also studies Java field, argument, and variable naming structures \cite{butler:2015}. Among other results, they identify noun phrases as the most common pattern for field, argument, and variable names. Verb phrases are the second most common. Further, they discuss phrase structures for boolean variables; finding an increase in verb phrases compared to non-boolean variables. 

The prior two papers by Butler and Singer focus on class names \cite{singer:2008, butler:2011}. Singer primarily identifies the noun phrase grammar pattern so that they can use it to match class names and study micro patterns (i.e., design patterns). Butler focuses more heavily on the class name grammar patterns, with the difference between his work and ours being the focus. Butler focuses on identifying how patterns in a class name repeat in super classes or interfaces, which limits the number of patterns they discuss. We focus on more types of identifiers (i.e., not just class names) and include a larger variation of class name patterns since we are not focusing on on pattern use in super classes/interfaces. In later work \cite{butler:2015}, Butler adds fields, arguments, and variables. Both papers observe the use of adjectives before nouns but do not discuss noun-adjuncts. The noun phrases in Butler's work make no distinction between nouns and noun-adjuncts; they are all annotated as nouns. We argue, and show, that there is a distinction between them; noun-adjuncts play an important supporting role in the comprehension of head-nouns within noun phrases and should be annotated to reflect this fact for future tools to leverage that metadata. Further, the latter paper \cite{butler:2015} discusses patterns at the phrasal level and does not show or discuss grammar patterns at the level of granularity that our work does. This allows us to show and study these patterns more specifically to understand the diversity in pattern structure as well as common, and divergent pattern structures.

Olney \cite{Olney2016} also compared taggers for accuracy on identifiers, but only on Java method names which were curated to remove ambiguous words (e.g., abbreviations). Gupta et al \cite{Gupta:2013} designed Posse; a part-of-speech tagger built to annotated source code. While this paper does not study grammar patterns explicitly, they do discuss some grammatical observations leveraged by Posse to help increase its accuracy on source code. They separate these into two categories: method name observations, and class/attribute name observations. For method names, they observe, among other things (we elide to save space): 1) Function names beginning with the base form of a verb sometimes lack the entity/object that is the target of the action. 2) Some function names have an implicit action (e.g., elementAt has an implicit \textit{get} action). For classes, attributes, they observe that noun phrases are typical but note that booleans are frequently exceptions. Our work is primarily complimentary to theirs, as we have shown weaknesses in Posse's tagging algorithm and discuss (in the next Section) some ways to solve these problems.

Binkley et al \cite{Binkley:2011} study grammar patterns for attribute names in classes. They come up with four rules for how to write attribute names: 1) Non-boolean field names should not contain a present tense verb, 2) field names should never only be a verb, 3) field names should never only be an adjective, and 4) boolean field names should contain a third person form of the verb “to be” or the auxiliary verb “should”. This work focuses primarily on attributes; we deal with a larger variety of identifiers and discuss a larger range of grammar patterns. Our data confirms some of Binkley's findings, particularly in the use of verbs in boolean field names.

\section{Discussion} \label{Discussion+conclusion}
This study has implications for both the study of identifier names and the improvement of part-of-speech tagging techniques applied to identifiers. Below, we discuss how our results augment researchers' understanding of identifier names and how these results can be applied to improve part-of-speech taggers for identifiers.

As a general takeaway, \textbf{the implementation is a very important resource for annotating identifier names with the correct part-of-speech}, as we will discuss more in the takeaways below. Words in code change meaning based on context just like in English, but unlike English, that context is not always in natural language; sometimes, the context is the behavior specified by code. One example from above is the identifier \textit{do forward}, which we gave a grammar pattern of \textit{V V} due to the fact that it is a method that redirects an entity which is not present in the identifier name (i.e., the user) when it is called. However, if we change the method's implementation so that instead of being used as a synonym for \textit{redirect}, \textit{forward} refers to (for example) moving a video game sprite forward on a screen, then we might give it the grammar pattern \textit{V VM} (though \textit{move} might be a better verb than \textit{do} in this case). This also means that domain knowledge and code analysis may be critical resources to correct part-of-speech tagging in some situations.

\subsection{Takeaways from RQ1} 
\textbf{Noun-adjuncts are ubiquitous in identifier names}: Noun phrases are the most common grammar pattern outside of functions, where verb phrases are more common. However, unlike noun and verb phrases in English, the adjectives in these identifiers tend to be \textit{noun adjuncts} (i.e., noun modifiers); nouns which behave like adjectives. This fact highlights how developers use noun-words within the domain of the software, or computer science in general, as adjectives to modify the meaning of the entity represented by the head-noun, which is typically the last (i.e., rightmost) word in the identifier. The ubiquity of noun-adjuncts has important implications for part-of-speech techniques, especially with respect to unknown word handling, as shown by Swum's accuracy. Noun-adjunct should be the default unknown word annotation for certain contexts-- specifically words at the beginning or middle of declaration-statement, attribute, or parameter variables that have non-collection and non-boolean types. With the exception of Swum, none of the part-of-speech taggers we studied are effective at detecting noun adjuncts or, as a consequence, head-nouns.

Likewise, other tools which use part-of-speech to perform analysis should at least be aware of the relationship between noun modifiers and head-nouns, as well as their ubiquity, especially when leveraging a tagger which does not accurately recognize them. Otherwise, they will potentially misunderstand which word is the head-noun; it is important to distinguish the head-noun from adjectives and noun-adjuncts which specify it. The ubiquity of noun-adjuncts within noun phrases has also not been discussed or quantitatively confirmed in research prior to this paper. Prior work has discussed noun phrases and compound words. Butler \cite{butler:2011, butler:2015} and Deissenboeck and Pizka \cite{Deissenbock:2005} observe the use of adjectives before nouns but do not discuss noun-adjuncts. The noun phrases in both studies make no distinction between nouns and noun-adjuncts; they are all annotated as nouns. However, Deissenboeck and Pizka do mention the use of compound words in programming which follow a \textit{modifier head-noun} pattern similar to noun modifiers. We argue, and show, that there is a distinction between nouns and noun-adjuncts; noun-adjuncts play an important role in the comprehension of head-nouns and should be annotated to reflect this role such that the difference between them is plainly obvious to tools and researchers that leverage POS annotations. Gupta \cite{Gupta:2013}, who created Posse, made the same noun-adjunct observation as we do (i.e., they annotate using NM), but give no data or discussion about their frequency.

\textbf{Several identifier characteristics have an effect on part-of-speech annotation}: Function identifiers are more likely to contain a verb and be represented by a verb phrase. Attribute, parameter, and declaration-statement identifiers typically have singular noun-phrase grammar patterns \textbf{unless} they have a type that is a collection or Boolean. Collection type identifiers have somewhat increased probability of using a plural head-noun, causing them to be plural noun phrases. This observation does not hold for function identifiers with collection return types because plural function names are more likely to be a reference to the multiplicity of the entity (e.g., an object) being operated on by the function. Boolean type identifiers are more likely to contain a verb. These patterns are opportunities for researchers to recommend when developers should be using plurals and verbs in their identifier names. These could initially be simple reminders; asking developers to consider using a verb or a plural and giving them the option of ignoring or accepting the recommendation. As research more firmly grasps the situations in which these practices should be followed, suggestions can be strengthened. This would not be very invasive, and might help increase consistency in naming practices. The characteristic of Boolean-type identifiers containing verbs has been observed before by H{\o}st and {\O}stvold, Binkley, and Gupta \cite{Binkley:2011, Gupta:2013, Host:2009} and quantified by Butler \cite{butler:2015}; our work helps confirm their results by considering a larger number of identifiers from a broader range of categories (function, attribute names, etc).

\subsection{Takeaways from RQ2}
\textbf{Part-of-speech taggers still require significant improvements to be effective on identifiers, and may be augmented by combining their individual strengths. Preambles are a significant problem for all taggers}: Our results show that even the best tagger for identifiers has an accuracy of between 50.2\% and 67.8\%. The other two taggers had much lower accuracy at the grammar pattern level. At the part-of-speech tag level, Stanford was the most accurate tagger; able to annotate not only a larger set of part-of-speech tags than Swum or Posse, but also more accurately than both in many cases. The contrast in performance between the taggers at different levels (i.e., grammar-pattern level and part-of-speech level) indicates that these approaches are likely complimentary; their combined output can help us find the correct grammar pattern. Some of these are straightforward. Noun plurals, for example, are not annotated by Swum or Posse. Stanford has 71.85\% accuracy on them and could inform us when Swum or Posse has missed a plural. Stanford is also the most effective at annotating digits, having agreed with them human annotations 100\% of the time; the second-best tagger for digits is Swum at 18.52\% accuracy. Stanford is also the only tagger that annotates conjunctions somewhat correctly (although, there were not many in the data set), while Posse and Stanford are more accurate at detecting prepositions than Swum; Swum has 29.79\% accuracy on prepositions while Posse and Stanford have 62.77\% and 90.43\% accuracy respectively. On the other end, Swum is far more effective than Stanford and Posse at correctly annotating noun modifiers (noun-adjuncts) with an accuracy of 94.01\% versus Stanford's 15.71\% and Posse's 23.25\%. 

All in all, the strengths of one tagger are the weaknesses of the others. We can take advantage of the strengths and weaknesses of each tagger to produce better part-of-speech tagging annotations. Increasing overall accuracy on plural words and digits by combining Swum and Stanford's output is very simple and will cause immediate improvements in part-of-speech annotations. Increasing accuracy on other part-of-speech annotations will require more sophistication, but our data indicates that ensemble tagging \cite{DellOrletta2009EnsembleSF} may be an effective approach. Further, all taggers have significant problems detecting preambles due to fact that preamble detection requires taggers to recognize naming conventions and domain information. Future research should focus on building naming convention recognition into taggers more effectively and finding ways to determine when a preamble is being used as a namespace. Frequency and position within the identifier might be useful in this regard, but will likely lead to false positives. It may be that domain terms and abbreviations will need to be identified as a configuration step before tagging. To our knowledge, this is the first formal comparison of tagger output and accuracy which has pinpointed specific areas of tagger synergy for identifiers in source code across multiple types of identifiers. Olney performed a comparison of taggers, but only on method names \cite{Olney2016} and they do not give an in-depth discussion of individual tagger weaknesses on either individual parts of speech or on grammar patterns.

\textbf{Certain identifier configurations can improve the output of off-the-shelf taggers:} Stanford+I is a technique where one adds an \textit{I} before a method name to help Stanford identify verbs more effectively. Our results confirm this-- prepending \textit{I} before method names increased the accuracy of Stanford by around 3\%. In addition, we tried two ways of feeding data to Stanford. Typically, Stanford reads sentences and paragraphs from a text file. We tried both feeding 1 identifier per file (i.e., 267 files, each with 1 identifier) into Stanford and multiple identifiers in a single file (i.e., 1 file with 267 identifiers) separated by a period and a new line. Stanford's grammar pattern level accuracy did not change under either of these configurations, although it did make different tagging decisions in a small minority of identifiers. As these changes did not affect its grammar pattern level accuracy, and made minor changes overall, we did not report numbers. However, we do note that the way identifiers are fed to Stanford will have some small impact on the part of speech annotation it uses for some words. The Stanford+I technique has been used in prior research \cite{Olney2016}. In this paper, we confirm that it improves Stanford's tagging on function names.

\textbf{Certain verb conjugations are frequently used as adjectives in certain types of identifiers:} Past tense verb (VBD), present participle verb (VBG), and past participle verb (VBN) are all used as adjectives in many situations within our data set. For example, \textit{sortedIndicesBuf}, \textit{waitingList}, and \textit{adjustedGradient} where \textit{sorted} is a past tense verb (VBD), \textit{waiting} is a present participle (VBG), and \textit{adjusted} is a past participle verb (VBN). While these are all verbs, they are used to describe the head-noun in their respective identifiers and should therefore be annotated using \textit{NM}. One particular situation where this happened a lot is when there was a non-function variable referencing a domain of functions. For example, \textit{get Protocol Method} is a parameter which holds a reference to a method, which can be one of many different methods passed in through the parameter. In other words, \textit{get protocol method} is a noun phrase with the grammar pattern \textit{NM NM N}, where the words \textit{get} and \textit{protocol} describe the type of method represented by this parameter instead of the action of a get-method. Where \textit{get} would typically be a verb, it is instead a noun modifier. We find that these verb-adjectives are typically found in attribute, parameter, and declaration-statement identifiers, meaning that verbs in these situations need to be more highly scrutinized than when they are found in functions. Detecting when an identifier represents a function, event, or generally any function callback could significantly help detect verb-adjectives. In addition, verbs that are of one of the three forms above and are found in attributes, declaration-statements, or parameters are more likely to be noun modifiers than they are to be verbs, as shown by the fact that Stanford's output improved when we interpreted them as noun modifiers instead of verbs (Table \ref{tab:accuracytable}). This can be directly applied to improve off-the-shelf tagger accuracy and, as a result, any tagging which relies on the output of multiple taggers since Swum and Posse, for example, do not use these conjugated verb annotations and are less able to determine when a verb should be an adjective. This result also means that part-of-speech taggers for source code should support these conjugated verb forms. 

Posse and Swum's developers both recognized the difficulty of tagging verb conjugations \cite{Gupta:2013,HillSwum:2010}; H{\o}st and {\O}stvold also observes this for method names \cite{Host:2009}. We provide data on this problem that the prior three papers do not include. Specifically our data set includes verb conjugations in a larger sample of manually validated grammar patterns that can be used for further analysis and tagger training; we discuss one way to help determine how to interpret verb conjugations in a source code context (i.e., use verb for function and noun-adjunct for other categories); and we show that this interpretation does matter, since Stanford's accuracy varies with how we interpret verb conjugations.

\subsection{Takeaways from RQ3}
\textbf{Many grammar patterns in our data set were infrequent, but still appeared to represent specific forms of behavior:} Grammar patterns that contain prepositional phrases and determiners may appear more frequently in software following specific design patterns or software architectures. For example, prepositional phrase grammar patterns like \textit{P NM* N} or \textit{V P} correlate with event-driven-programming functionality with identifiers such as \textit{onclick()} and \textit{onKeypress()}; conversion functionality such as \textit{ToString()} and \textit{ToInt()}; or common data analysis operations such as \textit{order by} and \textit{group by}. The same holds true for patterns containing determiners. Identifiers such as \textit{all action roots}, \textit{all open indices}, and \textit{all invocation matchers} should be more closely studied to understand more about their usage contexts. While prepositional phrase and determiner patterns were not common in our data set, they were used in several systems independently. Some of these patterns are more typical in certain domains. Prepositional phrase grammar patterns are used in event-driven programming very frequently, for example. The patterns presented in RQ3 represent future directions for research; there are not enough examples of the patterns we presented in this research question to draw strong conclusions, but their existence is evidence that these grammar patterns are common in other contexts (e.g., different architecture and design patterns). We argue that studying grammar patterns in these other contexts may result in more examples or even new patterns. These domain-specific patterns would then be very useful for suggesting and appraising identifier names within those contexts.
\subsection{Takeaways from RQ4}
\textbf{The most common C/C++ and Java grammar patterns are very similar, but they differ in their usage of full words and abbreviations in identifiers:} We observed a difference in the rate at which C/C++ and Java use abbreviations in their identifiers. This difference may make it more difficult for part-of-speech taggers, and other natural language approaches, to analyze C/C++ systems compared to Java systems. In addition, our results suggest that if we only control for identifier category and programming language, then use of Java or C/C++ does not have a significant impact on common identifier grammar patterns for systems written in these languages. However, there are a number of other confounding factors which future research should explore in order to more definitively ascertain the difference (or lack thereof) between grammar patterns found in different programming languages. Some factors which we recommend future research control for are: architecture patterns, design patterns, problem domain, programming paradigm, and modernity; some of our data (e.g., RQ3) suggests that these may have an influence on the grammar patterns identifiers follow.

\section{Conclusions and Future Work} \label{Conclusion}

We have identified and discussed a frequent, specific, and diverse, set of grammar patterns which developers use to express program semantics and behavior. These patterns are obtained directly from observations made on source code and provide researchers with a glimpse into how different grammar patterns are used to express different forms of program semantics. We do not argue that we have discovered all grammar patterns. On the contrary, our discussion from RQ3 suggests that there are more to be discovered. The grammar patterns presented in this study are a starting point for the creation of a taxonomy which should be curated over time and through further research. The dataset that we used to perform our comparison contains a fine grain set of grammar patterns from the systems we studied, which is a more granular view of these patterns than presented in prior work. This dataset is available on our webpage \footnote{https://scanl.org/} to help improve POS taggers for software and encourage replication.

Our direct future work includes: 1) improving POS taggers for source code by leveraging ensemble tagging \cite{DellOrletta2009EnsembleSF}, 2) using grammar patterns and static analysis models \cite{Dragan:2006, Gil:2019,Alsuhaibani:2015} to appraise identifier names, and 3) studying grammar patterns in specific design/architectural patterns and on test code. The results from this work will be particularly useful for improving the analysis of words, and their relationship to other nearby words, in the lexicon of a given software system. In particular, our observations will allow for better recognition of head-nouns (or lack thereof), verb-direct object pairs, and phrases used by code search and modeling tools like Swum and Sando \cite{shepherdsando, HillSwum:2010} and provide support for AI/ML techniques which recommend names \cite{Allamanis:2015, Liu:2019}; allowing them to more effectively choose words, even neologisms, that fit an appropriate pattern. The grammar patterns we have identified, and future extension to the collection, will provide greater understanding of the role different patterns play in comprehension and generation of identifier names.

% \begin{table}[]
% \centering
% \caption{Examples of frequent patterns from tool annotated set that were not in manual set}
% \label{tab:toolannotatedexampletable}
% \begin{tabular}{llll}
% \hline
% \multicolumn{4}{c}{\textbf{Grammar Pattern: NM NM}} \\ \hline
% \multicolumn{1}{l|}{Identifier} & \multicolumn{1}{l|}{Posse} & \multicolumn{1}{l|}{Swum} & Stanford \\ \hline
% \multicolumn{1}{l|}{identity matrix} & \multicolumn{1}{l|}{N N} & \multicolumn{1}{l|}{N N} & NM NM \\ \hline
% \multicolumn{1}{l|}{extended class} & \multicolumn{1}{l|}{NM N} & \multicolumn{1}{l|}{V N} & NM NM \\ \hline
% \multicolumn{1}{l|}{search path} & \multicolumn{1}{l|}{N N} & \multicolumn{1}{l|}{V N} & NM NM \\ \hline
% \multicolumn{4}{c}{\textbf{Grammar Pattern: V NPL}} \\ \hline
% \multicolumn{1}{l|}{get predicates} & \multicolumn{1}{l|}{V N} & \multicolumn{1}{l|}{V N} & V NPL \\ \hline
% \multicolumn{1}{l|}{persist quantiles} & \multicolumn{1}{l|}{V N} & \multicolumn{1}{l|}{V N} & V NPL \\ \hline
% \multicolumn{1}{l|}{swap yields} & \multicolumn{1}{l|}{V N} & \multicolumn{1}{l|}{V N} & V NPL \\ \hline
% \end{tabular}
% \end{table}

\section{ACKNOWLEDGEMENT} \label{sec:acknowledgement}
This material is based upon work supported by the National Science Foundation under Grant No. 1850412.
\newpage
\bibliography{references}

\end{document}